# Interactive Virtual Games: Winners for Deep Cognitive Assessment

Dom CP Marticorena[1], Zeyu Lu[2], Chris Wissmann[1], Yash Agarwal[3], David Garrison[3], John M. Zempel[3], Dennis L. Barbour[1,2]

[1]Department of Biomedical Engineering, Washington University, 1 Brookings Drive, St. Louis, MO 63130
[2]Department of Computer Science and Engineering, Washington University, 1 Brookings Drive, St. Louis, MO 63130
[3]Department of Neurology, Washington University School of Medicine

# Abstract

Studies of human cognition often rely on brief, highly controlled tasks that emphasize group-level effects but poorly capture the rich variability within and between individuals. A suite of minigames built on the novel pixelDOPA platform was designed to overcome these limitations by embedding classic cognitive task paradigms in a 3D virtual interactive environment with continuous behavior logging. Four of the minigames explore constructs that overlap established NIH Toolbox tasks, including processing speed, rule shifting, inhibitory control and working memory. Across a clinical sample of 66 participants collected outside a controlled laboratory setting, large correlations ($r$ = 0.47–0.92) between the pixelDOPA tasks and NIH Toolbox counterparts were found, despite differences in stimuli and task structures. Process-informed metrics (e.g., gRT) substantially improved both task convergence and data quality. Test-retest analyses revealed high reliability (ICC = 0.71–0.85) for all minigames. Beyond endpoint metrics, movement and gaze trajectories revealed stable, idiosyncratic profiles of gameplay strategy, with unsupervised clustering differentiating participants by their navigational and viewing behaviors. These trajectory-based features showed lower within-person variability than between-person variability, facilitating participant identification across repeated sessions. Game-based tasks can therefore retain the psychometric rigor of standard cognitive assessments while providing new insights into dynamic individual-specific behaviors. By leveraging a highly engaging, fully customizable game engine, comprehensive behavioral tracking can boost the power to detect individual differences without sacrificing group-level inference. This possibility reveals a path toward cognitive measures that are both psychometrically robust and deployable in less-than-ideal settings, while capturing richer behavioral data than traditional paradigms.

# Introduction

Understanding individual differences in cognition is fundamental to psychology, neuroscience, and related fields, but most current assessment methods are optimized for detecting group-level effects. They rely on brief, highly controlled tasks that aim to minimize noise (Cronbach, 1957). While this procedure reduces extraneous variability, it also means that a substantial amount of informative individual variation is filtered out in the process (Fraley & Roberts, 2005; Hedge et al., 2018; Rouder et al., 2023). With tasks designed primarily for simple between-condition contrasts, the unmodeled variance associated with differences between people can overwhelm the effects of interest (Rouder et al., 2023), underscoring the need for individual difference methods that preserve and leverage how cognition varies across people and contexts.

Recent work highlights the value of capturing more informative data to reveal genuine differences in how people perceive, decide, and act (Hedge et al., 2022; Linari et al., 2022). Two broad strategies have emerged: first, increasing the number of trials per participant (i.e. trial count) to stabilize estimates of target constructs. Second, some researchers have prioritized expanding what we term “trial depth”: instead of measuring endpoint accuracy or final response time, they record continuous, within-trial behavior. This process information includes computer mouse trajectories (Dale & Duran, 2011; Smith et al., 2022) and continuous gaze data (de Sardenberg Schmid & Hardiess, 2024), among others. Such information is used to disentangle underlying cognitive processes, such as distinguishing between stimulus encoding versus response selection stages, automatic versus controlled processing, and temporal dynamics of conflict monitoring (Brunetti et al., 2019; Scherbaum et al., 2018). This approach can also uncover variability otherwise lost in aggregate measures (Koenig-Robert et al., 2024).

Yet both approaches face the challenge of fatigue: prolonged testing can sap motivation, especially in the highly controlled environments often needed for added instrumentation (Shute, 2009). Emerging gamification techniques (e.g., applying points, badges, etc., to induce engagement in otherwise unengaging tasks) are often superficial and insufficient for sustaining engagement over dozens or hundreds of trials (Hamari et al., 2014; Sailer et al., 2017). Truly gameful designs require participants to immerse themselves in a goal-driven environment that fosters extended, continuous interaction (Deterding et al., 2011).

One especially promising avenue for achieving such challenge-based immersion is 3D virtual interactive environments, which approximate real-world navigation and decision-making more closely than 2D or single-button tasks (Parsons, 2015; van der Kuil et al., 2018). Although pencil-and-paper tasks were once the neuropsychological gold standard, current technology enables the kind of richly contextualized data collection that 3D virtual interactive environments make possible on a large scale. They provide realistic movement, multi-axis controls, and dynamically evolving scenarios to maintain participant interest (Chicchi Giglioli et al., 2021; Sudár & Csapó, 2024). Interfaces for controlling behavior in 3D virtual interactive environments are often carefully designed in an intuitive manner (Limperos et al., 2011; McEwan et al., 2012; Pinelle et al., 2008), which reduces frustration and participant dropout (Low et al., 2017; van der Kuil et al., 2018). These features enable more trials and richer data without the motivational decline characteristic of repetitive lab-based paradigms (Deterding et al., 2011). These design features suggest that 3D virtual interactive environments offer three complementary advantages for cognitive assessment: 1) richer, continuous behavioral data for inferring cognitive processes, 2) sustained engagement that enables longer and deeper sampling, and 3) increased ecological validity, potentially arising from the combination of the first two. Despite this potential, few tools fully harness 3D virtual interactivity for continuous, process-level cognitive assessment while maintaining sustained engagement at scale, which is a gap our current work aims to fill.

To address this gap, we introduce pixelDOPA, a 3D virtual interactive environment with minigames designed to target classical executive functions including rule shifting, inhibitory control, processing speed and working memory, designed for extended play. These constructs were selected because they are among the most assessed domains in cognitive testing, enabling validation against well-established standardized measures. The platform is built atop the Minecraft codebase, chosen for its flexible ecosystem, active modding community, and intuitive control interface (Johnson et al., 2016; Peters et al., 2021). The open software structure naturally accommodates continuous data logging, enabling researchers to capture richer behavioral metrics (e.g., movement, orientation, actions) without intruding on gameplay. Our primary aim with the current work is to demonstrate that analogues of classical executive function tasks defined in a 3D virtual interactive environment can meet the psychometric standards of gold-standard assessments (e.g., NIH Toolbox). Upon establishing this finding, we then explore whether the resulting expanded trial count and trial depth afforded by this novel platform can uncover stable individual differences even for these reductionist tasks originally designed to resolve group differences.

We report an initial validation of pixelDOPA in a hospital neurology unit—an environment characterized by frequent clinical interruptions, variable medication schedules and diverse participant backgrounds. This setting was selected because it offers a diverse population varying in age, socioeconomic status and cognitive abilities. Extended hospital stays enabled repeated testing sessions, both for assessing test-retest reliability and for derisking a longer-term study of individual cognitive fluctuation in this population. We assessed whether pixelDOPA tasks correlate with NIH Toolbox measures and exhibit test-retest reliability despite these uncontrolled conditions, whether process-level metrics such as avatar-gaze-based responses could recover trials otherwise discarded under classical criteria, and whether task behaviors reveal stable participant-specific differences. Despite these uncontrolled conditions, pixelDOPA tasks correlated strongly with NIH Toolbox measures and exhibited high test-retest reliability. We further found that continuous process-level metrics, such as avatar-gaze-based response times, can rescue trials otherwise discarded based on classical endpoint metrics. Finally, we determined that unique movement trajectories reveal participant-specific strategies and executions that remain stable across sessions.

Our results indicate that minigames implemented in a 3D virtual interactive environment can produce richer, process-level behavioral data (e.g., avatar gaze and movement) while achieving strong convergence with standardized NIH Toolbox measures and high test-retest reliability, even under uncontrolled clinical conditions. These features allow pixelDOPA to capture how individuals navigate and attend to tasks, beyond endpoint accuracy or response time alone. Having established strong convergence and reliability, the present work positions expanded trial depth and continuous behavioral logging as promising tools for charactering individual-specific task behaviors. Evaluating links between such task profiles and real-world behaviors or outcomes remains an important goal for future hypothesis-driven studies.

# Methods

## Cognitive Test Batteries

### Minigame Battery

A suite of minigames (gamefully designed simple tasks) was developed to emulate established cognitive testing paradigms within a 3D virtual interactive environment. Three of the minigames (Nether Knight, Door Decipher and Barnyard Blast) were designed as alternative-forced-choice (AFC) tasks that emphasize speeded responses and offer close points of comparison to classic executive function assessment tasks. Screenshots of these minigames as the participant experiences them are provided in **Figure 1**. While accommodating participant preference for avatar control and task completion strategy, we incorporated structural limits and guided progression to ensure that participants do not wander off task or otherwise perform activities uninformative to the questions at hand. For example, dynamic walls and barriers prevented movement into trial areas not yet active or already completed, and player inventories were restricted to only those items relevant to the current task.

Nether Knight (NK): Depicted in Figure 1a, participants enter a trial where two knight figures of distinct colors appear, along with a centrally located, colored word spelling out a different color that remains on screen for 5000 ms. Prior to beginning, participants are presented with instructions indicating whether they should focus on the stimulus's semantic content or its display color. Instructions must be remembered for the duration of the session. For example, in semantic-based trials, a stimulus with the word “BLUE” displayed with red letters would require interacting with the knight whose armor is blue, while for color-based trials, the same stimulus would require interacting with the red-armored knight. Participants must interact with the correct knight by walking up to it and selecting it with a mouse click. After responding, feedback (correct or incorrect) is displayed for 1500 ms, and participants proceed to the next trial area. A complete session contains a total of 40 trials.

Door Decipher (DD): In each trial, participants receive a “key card” showing a specific color (red/blue/green/yellow), shape (circle/square/triangle/plus), and quantity (1/2/3/4). Shown in Figure 1b, four doors are arrayed before the participant, each topped with an overhead card that also shows color, shape, and quantity attributes. Participants must identify the currently relevant attribute (color, shape, or quantity) and enter the door with the single matching attribute that corresponds to the given key card. Every seven trials, the relevant dimension changes without warning, requiring participants to determine the new rule and adjust their choices accordingly. After responding, feedback (correct or incorrect) is displayed for 1500 ms, and participants proceed to the next trial area. A complete session contains a total of 42 trials.

Barnyard Blast (BB): At the start of each trial, five animals (either all cows or all pigs) appear in a row, with the central animal serving as the target stimulus, as shown in Figure 1c. Participants are instructed to respond according to the direction the central animal is facing while ignoring the flanking animals by aiming at and tagging a left or right target. If the animals are cows, the response direction matches the central cow's orientation; if the animals are pigs, the response direction is the opposite of the central pig's orientation. The task includes "congruent" trials (flankers facing the same direction as the center) and "incongruent" trials (flankers facing the opposite direction). Feedback (correct or incorrect) is displayed for 1500 ms before the next trial begins. A complete session contains a total of 48 trials, with 12 from each stimulus-congruence combination.

In addition to the timed AFC tasks above, we also designed minigames with different response mechanics. Rainbow Random is an adaptive task to estimate visuospatial working memory. It diverges from the AFC format by focusing on the accuracy of a reconstructed geometric structure rather than on rapid response selection.

Rainbow Random (RR): This minigame employs an adaptive, staircase-like design to assess participants' ability to reconstruct a 2D spatial arrangement of blocks composed of up to three colors. Depicted in **Figure 2**, each trial begins with a 5-second preparation phase, followed by a 60-second observation phase during which participants must study a randomly generated pattern. If desired, participants may exit the observation phase early as soon as their memorization is complete. In the subsequent 60-second build phase, they reconstruct the pattern by placing blocks to match both spatial arrangement and color. They may also exit the build phase early by submitting their reconstruction. After the build phase, a 5-second judging phase provides accuracy feedback and if they passed or failed, determined by whether reconstruction accuracy exceeds 80%. A session begins with a single block on the first trial and automatically adapts based on performance: successful replication increases the pattern size by one block for the next trial (to a maximum of 10), while unsuccessful attempts decrease it by a block. Trials continue until a participant fails twice at the same difficulty level, or until the maximum pattern size is replicated. As with all the other minigames, a tutorial session precedes the main task.

Different game mechanics were purposefully designed into each of these minigames to evaluate how moving an avatar versus performing precise mouse clicks might influence data quality and engagement. For instance, NK involves moving the character, adjusting the view, and clicking to select a target; DD requires no click-based selection but does involve navigating through the correct door by moving and adjusting one's view; and BB includes view adjustment and a click to select a small target, though there is no character movement. These design decisions were intended to reveal which gameplay elements in a 3D virtual interactive environment might be best suited for acquiring informative data for cognitive inference. Any minigame redesign to take advantage of empirically determined preferential game mechanics would be easily implementable given the flexibility of the software platform.

## Traditional Assessment Battery

Participants also completed selected tests from the NIH Toolbox Cognitive Battery (NIHT; 7+ version) (Weintraub et al., 2013; Zelazo et al., 2013). These tests consist of the following:

Pattern Comparison Processing Speed (PCPS): Participants view pairs of simple visual patterns and decide whether the two images are exactly the same or different. They complete as many same/different comparisons as possible within a fixed time period.

Dimensional Change Card Sort (DCCS): Participants see two target pictures that differ in attributes such as color and shape. They are asked to sort a series of test cards by matching one of these attributes. After a block of trials, the sorting rule switches so that they must sort by the alternate attribute.

Flanker Inhibitory Control and Attention (FICA): A central stimulus arrow is presented with additional arrows flanking it on either side. Participants indicate the direction of the central stimulus while ignoring the surrounding items, which sometimes point in the same direction as the central stimulus and sometimes in the opposite direction.

List Sorting Working Memory (LSWM): Participants view images of foods and animals, presented with both audio and text labels. In single-category trials, they arrange either food or animal items by size from smallest to largest. In dual-category trials, they must first sequence all food items by size, then separately sequence all animal items by size.

These standardized tests served as benchmarks for comparison with the minigames. Although each minigame was designed to emphasize a particular cognitive domain that aligns with a corresponding NIHT measure, none of them were created as exact, one-to-one replications.

NK was conceptualized as a speeded task akin to PCPS but includes a selective attention component (e.g., resolving text vs. color) for each trial. DD deliberately focuses on rule-shifting, sharing similar cognitive demands as the DCCS, even though the specific manifestation of rules and stimuli differ. BB parallels the FICA task but introduces additional “reverse” trials, thereby adding further inhibition demands absent in the standard NIHT version. Finally, RR targets working memory capacity similarly to LSWM but uses a visuospatial block arrangement rather than purely verbal list recall (**Table 1**).

These four minigames were purposefully paired with their closest NIHT counterparts, forming our predefined cross-platform task pairs, to ensure conceptual alignment with well-known cognitive measures while allowing for broader design flexibility and added cognitive demands in anticipation of sophisticated latent variable analysis (Kasumba et al., 2025). These design liberties were taken to preserve gamefulness while maintaining each task’s primary cognitive dimension, rather than replicating NIHT protocols exactly. Despite such differences in form, we expected the dominant constructs (rule shifting, inhibitory control, processing speed, and working memory) to drive performance comparably across both platforms. Although the minigame implementation platform includes other cognitive assessment minigames and continually adds more, these four were selected for the present validity and reliability study due to their clear overlap in cognitive constructs with established NIHT measures and therefore their ability to validate the potential of an easily scripted 3D virtual interactive environment to resolve individual cognitive function.

## pixelDOPA platform

All the minigames described above were designed and constructed in a unified software platform we call pixelDOPA (Digital Online Psychometric Assessment). pixelDOPA is an integrated digital assessment environment built on Minecraft’s underlying game mechanics engine. By leveraging the software’s flexibility for rapid prototyping and iteration, pixelDOPA features a modular minigame interface, which is a plug-and-play system for rapidly creating and flexibly deploying a wide variety of cognitive assessment minigames. Hosted on a protected server accessible to participants, pixelDOPA comprises a central minigame-selection lobby and the modular minigame environments described above. Participants join the lobby upon connecting to the server, then access available minigames by interacting with game elements marked with each minigame’s details. Participants are assigned unique nicknames to identify themselves to the software. Minigames are organized to enable both systematic administration of specific tasks and open-ended “as-desired” play to enhance overall engagement by promoting participant agency. The lobby includes a shop where participants can purchase real-life items using currency earned from completing minigames, which enhances engagement further. A dynamically updating leaderboard highlights the top ten participants by nickname with the highest number of total minigame completions.

Prior to beginning any minigame, participants must complete a general gameplay tutorial (accessed from within the lobby) to familiarize themselves with basic movement controls. Upon emerging within the tutorial area, participants are first instructed to move their avatars forward, backward, left, and right, as well as look up and down and rotate (i.e., adjust pitch and yaw), as shown in **Figure 3a**. They then practice several additional but essential controls, such as jumping, sprint-jumping, and crouching. Next, participants receive an item in their quick-access bar to practice selecting and using it to interact with an in-game entity (**Figure 3b**). They also complete simple tasks requiring them to place (**Figure 3c**) and break (**Figure 3d**) blocks, learning how to retrieve items in the process. Finally, the tutorial concludes with participants using a crossbow to tag targets, reinforcing point-and-click mechanics. Upon completion, participants are returned to the main lobby, where they can then freely choose which minigames to play.

## pixelLOG functionality

pixelLOG (Logging of Online Gameplay) is a specialized, highly customizable data-collection toolkit designed to capture fine-grained, player-level information for activities defined on pixelDOPA and other platforms (Lu & Barbour, 2026) (**Figure 4**). Inspired by earlier Minecraft-based reinforcement learning frameworks such as Microsoft Project Malmo for agentic AI (Johnson et al., 2016), pixelLOG expands these capabilities to support comprehensive, human-focused psychometric assessments. By providing configurable sampling rates and exhaustive behavioral tracking (including avatar position, orientation, target views and environmental interactions), pixelLOG yields a richly detailed, process-based representation of how individuals interact within the 3D virtual interactive environment.

Data collection in pixelLOG proceeds via two distinct polling frequencies. High-frequency polling (20 Hz) captures rapidly changing player states (for example, location and perspective), whereas lower-frequency sampling (≥1 Hz) records comparatively stable environmental attributes such as block configurations and static entities. In addition, pixelLOG integrates event-driven data capture by intercepting and logging meaningful player actions as discrete events. Both the actively polled state data and passively recorded event data are synchronized and time-stamped, then organized into a sequential log, constituting the primary component of the log file.

pixelLOG further collates trial-specific data across all pixelDOPA minigames, including (but not limited to) trial start and end timestamps, stimulus parameters and trial outcomes. These data are consolidated into a secondary trial summary component of the log file. This dual-structured format facilitates the extraction of trial-specific segments from the primary log sequence and enables precise reconstruction of the stimulus environment encountered by players during each trial. Such features are critical for efficient downstream analysis, visualization and interpretation.

# Data Collection

## Participants

This validation study was embedded within a larger study of cognition in epilepsy and related neurological disorders for children and young adults with a wide variety of gaming experience. Over a fourteen-month period, 539 individuals aged 5 and older were admitted to the St. Louis Children's Hospital EEG Monitoring Unit. Research staff determined that 395 of these individuals passed eligibility criteria (able to read and follow instructions). Of the eligible participants, 195 were approached, 48 declined to participate, and 67 individuals were assigned and consented into another experiment, leaving 80 individuals consented into this study. However, the data from the first 7 enrollees (who served as pilot participants during the initial data-collection setup) could not be processed for technical reasons. This left a final sample of 73 participants (ages 5 to 24 years; 48% female) for the pixelDOPA (pD) study. All procedures were approved by the Institutional Review Board at Washington University in St. Louis, and informed consent was obtained from all participants or their guardians.

Of the 73 consented participants, 3 ended their tasks or minigames early, yielding insufficient trials. This resulted in 70 individuals who contributed data to the pD study. Of these, 4 were excluded due to extreme outlier responses in one of their tasks or minigames. For Study 1 (convergent validity), 35 individuals each completed at least one predefined cross-platform task pair. For Study 2 (test-reliability), 52 individuals each completed at least one pD minigame on two occasions. Twenty-one individuals participated in both studies **(Figure 5)**. Due to the nature of the underlying scientific study, participants selected the order and amount of testing they pursued. This flexible ordering of minigames served two purposes: it provided practical design feedback for future development iterations, and it also accommodated the practical constraints of the inpatient setting. As a result of this flexibility, the unknown amount of time for research activities in between clinical testing, and the unpredictable total time of their hospital stay, not all the 66 final participants were logistically able to complete all possible tasks at least once before being discharged.

## Data Collection Protocol

### NIHT Assessments

All participants were invited to complete a subset of NIHT cognitive tasks administered in their hospital rooms on an iPad. Each task was presented separately, and while participants were encouraged to finish every task in the battery, some refused or skipped certain tasks due to fatigue, lack of interest, or logistical constraints. In total, each NIHT task required approximately 10 minutes. In accordance with standard NIHT operator guidelines, participants were supervised by trained research staff during testing and instructed to maintain a standardized hand position before each trial to minimize variability in response times. All NIHT data, including timestamps, accuracy, and response times, were automatically logged on the device and backed up to a secure server.

### pD Minigames

In addition to the NIHT assessments, participants were introduced to pD minigames. These were completed on a Dell XPS 13 laptop with remote access to the pD server. Upon joining the server for the first time, participants chose a non-identifying nickname and completed the tutorial (Figure 3) under the guidance of a research staff member. Each laptop was configured with uniform in-game sensitivity settings, and participants could choose between a standard keyboard + mouse configuration or a handheld controller for gameplay.

Exactly as was done with the NIHT tasks, no strict minimum number of minigames was required. Participants were encouraged to try each minigame, though completion rates varied due to factors including length of stay, medical procedures, and participant state. This variability in engagement, while introducing potential selection effects, reflects the real-world challenges of deploying cognitive assessments in clinical or home settings. To maintain engagement, in-game currency was awarded upon minigame completion and could be exchanged for small real-world prizes (e.g., stickers, toys). As with the NIHT tasks, a small proportion of participants opted out of certain minigames or discontinued due to boredom or clinical considerations. All minigame sessions were recorded using pixelLOG and backed up to a secure server.

#### Engagement and Monitoring

Participants completed all NIHT tasks and most pD minigames under supervision, with research staff providing only verbal encouragement and technical assistance (e.g., reconnecting to the server if accidentally disconnected). Staff avoided offering strategic guidance or task-specific hints to preserve the integrity of the data. Some participants, particularly those with longer hospital stays, completed multiple rounds of assessments and minigames over several days, providing the opportunity to evaluate test reliability. In cases where laptops were left with participants overnight, simple on-screen instructions were provided for accessing the pD server and resuming tasks while unsupervised.

## Data Analysis

### Data Cleaning and Outlier Exclusion

Data were collected in a clinical setting where distractions (e.g., staff visits, medication schedules, emotional strain, family presence) could compromise participants' task performance.

To mitigate potential confounding effects of inattention and extremely long response latencies, a multi-step procedure was prospectively adopted for data analysis (Leys et al., 2013; Ratcliff, 1993; Whelan, 2008). First, trials whose endpoint time exceeded a predetermined maximum cutoff of 10 seconds were excluded (Ratcliff & Tuerlinckx, 2002). If more than 25% of a session's trials surpassed that cutoff, the entire session was removed from further consideration. Remaining data were then evaluated via a log-median absolute deviation (log-MAD) threshold to exclude remaining ultra-long RTs in highly skewed distributions (Leys et al., 2013; Whelan, 2008). Finally, an optional participant-level outlier step removed values if they exceeded a 3.5-MAD ratio, ensuring that only the most extreme data points for each measure were flagged while preserving the rest of the participant's data.

## Parameter Creation

### Response Times

After filtering, mean RTs were computed for each participant in both the NIHT tasks (PCPS, DCCS, and FICA) and the analogous pD minigames (NK, DD, and BB). These means served as summary measures for correlation analyses and descriptive statistics.

### Gaze Response Times

For each trial across all minigames, avatar-based gaze response time (gRT) was measured as the earliest point at which participants fixated their avatar crosshair on their ultimately selected target without subsequent deviation. At every game tick, pixelLOG recorded the knight, door or block that participants were looking at or aiming toward. For each trial, the choice that participants ultimately selected (the knight they selected in NK, the door they entered in DD or the block they tagged in BB was identified. The resultant log was then scanned to find the earliest stable alignment with that same entity or location, and this timestamp was converted into seconds from the onset of the trial. This gRT served as an alternative, objective, process-informed endpoint to the more conventional final RT or button press, offering a view of when participants first detectably committed to their choice for the trial.

### Psychometric Thresholds

Threshold-seeking tasks (RR and LSWM) were analyzed using a two-parameter logistic function, where $\psi_\theta$ denotes the difficulty level at 50% success (i.e., the psychometric threshold) and $\psi_\sigma$ represents psychometric spread. Fits were performed using the Levenberg–Marquardt algorithm; if initial fits failed to converge, $\psi_\theta$ was re-initialized near the empirically estimated 50% accuracy point. To avoid unbounded estimates, $\psi_\theta$ was capped at 0.5 above the maximum difficulty, for instance, 10.5 in RR and 7.5 in LSWM. Fit quality was evaluated by the root-mean-square error (RMSE) of the residuals.

## Correlational Analyses

Following parameter extraction, three sets of values were available for each participant:

1. **Mean RTs** for each AFC task,
2. **Mean gRTs** for each AFC task,
3. **Psychometric thresholds** ($\psi_\theta$) for the threshold-seeking tasks.

Pearson correlations were then computed to examine relationships between 1) NIHT tasks and analogous pD minigames (e.g., FICA and BB) and 2) each pair of threshold-seeking tasks (e.g., LSWM and RR). Bayes factors were used to quantify the strength of evidence for each correlation. Participants missing data in one or both tasks of a given pairing were excluded from that specific correlation. All cross-platform comparisons were pre-specified based on construct alignment (**Table 1**) and tested as independent, confirmatory hypotheses; family-wise error correction was therefore not applied (Rothman, 1990).

All statistical relationships between different tasks were assessed using Pearson's correlation coefficient (*r*). Test-retest reliability was assessed using a two-way random effects, absolute agreement, single-rater/measurement intraclass correlation coefficient (ICC(2,1)). Statistical significance for Pearson correlations was determined using two-tailed *t*-tests against the null hypothesis of no correlation ($r = 0$), while ICC statistical significance was assessed using *F*-tests comparing within- and between-subject variance. Bayesian analysis was performed using Bayes Factors ($BF_{10}$) calculated with the Jeffreys-Zellner-Siow (JZS) prior (Bayarri & García-Donato, 2008), which places a Cauchy distribution (scale = $1/\sqrt{2}$) on the standardized effect size and an uninformative prior on the variance. Statistical significance for the identifiability analysis was determined by performing an equal-weight permutation test, with null distribution of the mean per-participant Jensen-Shannon identifiability scores formed by randomly reassigning session labels among participants over 100,000 iterations.

## NK Trajectory Analysis

To determine if more detailed analysis of data from traditional group-designed tasks might resolve individual differences, we conducted an exploratory evaluation of intermediate process data from the Nether Knight minigame. Full time-series movement and viewing angles captured how participants moved and oriented themselves toward targets. Each trajectory consisted of $T$ samples (up to a fixed maximum duration), reflecting the participant's lateral position ($x$), forward progress ($z$), angular difference between the avatar gaze and target (target heading), and the sine and cosine of that heading. All participants who contributed NK sessions from Study 1 or Study 2 were included ($n = 55$ participants, 330 sessions, 11,532 trials). Trials with fewer than two valid samples or with incomplete data were removed. To make the mirrored trajectories amenable for analysis, each left-selection trial was reflected about the $z$-axis so that all final target positions became effectively right-sided. This process consolidated all trial paths into a shared reference frame, simplifying comparisons across trials and between participants.

To enable clustering across trials of varying lengths, each trajectory was 1) resampled to a fixed length (e.g., 120 samples) via linear interpolation, and 2) summarized by a set of 27 extended features (e.g., minimum and maximum positional values, net distance traveled, time indices of peak heading deviations). These features were designed to capture both the shape (direct vs. curved) of the movement path and any distinct orientation patterns (e.g., deviating away from the target before turning back).

After scaling the 27-dimensional feature vectors, we projected them into a 2-dimensional latent space via UMAP (McInnes et al., 2020). The algorithm used a neighborhood size of 10, a minimum distance of 0.05, and retained the default Euclidean metric. We then performed DBSCAN (Ester et al., 1996) on the resulting points (eps of 0.30, minimum samples of 20) to identify cohesive clusters in the embedded space. The identifiability result ($p$ < 0.00001, $d$ = 7.6) far exceeds significance thresholds, suggesting results are robust to reasonable parameter variations. The small number of trials assigned to the “noise” cluster were excluded from further interpretation.

The remaining trials naturally fell into 4 dominant clusters from the DBSCAN procedure, which automatically determines the number of clusters and cluster assignment. Visual inspection of the major features distinguishing the resulting 4 categories of trajectories revealed that motion path and gaze direction differed the most between clusters based on this analysis. Each trial could be categorized as “direct” vs. “indirect” (based on whether the trajectory path deviated away from the target by more than a small lateral threshold, and “gaze” vs. “no-gaze” (based on whether participants’ recorded gaze fixated on the knight before the final selection). This procedure therefore ultimately resulted in four subtypes of trials: Direct + Gaze (DG), Indirect + Gaze (IG), Direct + No-gaze (DN), Indirect + No-gaze (IN).

To examine how each participant’s distribution of trial-types might differ (or remain consistent) across repeated sessions, we constructed a personal profile of cluster usage by computing the percentage of that participant’s trials in each of the four categories (DG, IG, DN, IN). For this analysis, we selected only participants with 2 or more NK sessions (N=42, 317 sessions, 11,138 trials). We then calculated the Jensen-Shannon distance (JSD) between all pairs of session profiles, comparing within-person sessions to between-person sessions.

Finally, we visualized whether these trajectory-based profiles could identify a particular participant from all the others in the study cohort. For each session, we ranked all participants' average profiles by ascending JSD from the session's participant profile. We then assigned a weighted "vote" of 1.0 to the top-ranked participant, plus partial votes (0.5 or 0.25) if the true participant appeared second or third in that ranking. Summing and normalizing these votes by row yielded a confusion matrix whose diagonal entries reflect how frequently the true participant was recognized by their movement trajectory pattern.

# Results

## Process-Informed Metrics

In addition to classical endpoint response times (RTs), we analyzed avatar gaze response times (gRTs), defined as the earliest stable avatar fixation on the ultimately selected target, as a process-informed measure of decision commitment.

A total of 2,892 trials were collected from the three pD AFC minigames (NK, DD, BB) and 2,995 trials were collected from the corresponding NIHT tasks (PCPS, DCCS, and FICA) in each participant's first valid session. Across all sessions including optional repeats, these totals were 6,018 and 3,185, respectively (Supplementary Materials, Section 10). NIHT tasks exhibited low exclusion rates, ranging from 1.7% in FICA to 2.0% in PCPS.

**Table 2** shows that under classical RT criteria, BB excluded the highest proportion of trials at 8.7%, while DD and NK excluded 7.6% and 5.3% of trials, respectively. Switching to gaze-based RT (gRT) reduced exclusion in two of the pD tasks: DD dropped from 7.6% to 5.5%, and BB from 8.7% to 7.1%. Across all sessions including optional repeats (Supplementary Materials, Section 10), the benefit of gRT was more pronounced, with BB exclusion dropping from 26% to 8.5% and DD from 17% to 13%. This procedure salvaged 23 DD trials and 24 BB trials that would otherwise have been excluded. However, NK showed a net loss of 14 trials under gRT criteria, reflecting stricter fixation requirements for that task. Although NK's exclusion rate was slightly higher for gRT than RT (6.6% vs. 5.3%), DD and BB both showed reduced trial exclusion under gRT criteria. All three tasks benefited in convergence with their NIHT counterparts, with NK showing the largest gain. Because first fixation is less influenced by post-decision artifacts or errors in mouse input and screen interactions, it retains trials that classical RT might discard based on anomalous or protracted finishing times.

## Study 1: Convergent Validity

Each pD minigame was paired with the NIHT task expected to reflect the most similar cognitive construct: NK vs. PCPS (both emphasizing rapid responding), DD vs. DCCS (rule shifting), and BB vs. FICA (inhibitory control and selective attention).

**Figure 6** illustrates the correlations between each pD minigame and its corresponding NIHT measure when examining the classical endpoint of mean RTs. Each plot symbol is one participant's first valid session for each task. NK and PCPS showed a large positive association ($r$ = 0.53, $p$ = 0.011, $BF_{10}$ = 5.7, $n$ = 22, RMSE = 0.023). DD correlated more strongly with DCCS ($r$ = 0.69, $p$ = $3.1 \times 10^{-4}$, $BF_{10}$ = 120, $n$ = 23, RMSE = 0.034), and BB correlated somewhat less strongly with FICA ($r$ = 0.47, $p$ = 0.028, BF = 2.5, $n$ = 22, RMSE = 0.048). These large positive correlations under such variable testing conditions support the hypothesis that each pD minigame engages overlapping cognitive constructs with its NIHT counterpart.

To examine whether earlier decisional processes might reflect better predictors of cognitive performance, we computed process-informed endpoints (i.e., gRTs) and compared them with classical RT-based correlations. As shown in **Figure 7**, the relation between BB and FICA remained similar, decreasing slightly from $r$ = 0.47 to $r$ = 0.42 ($p$ = 0.045, BF = 1.7, $n$ = 23, RMSE = 0.032). DD and DCCS showed a moderate increase, from $r$ = 0.69 to $r$ = 0.77 ($p$ = $1.7 \times 10^{-5}$, BF = $1.5 \times 10^{3}$, $n$ = 23, RMSE = 0.026). However, for NK and PCPS, gRT substantially improved metrics across the board, increasing from $r$ = 0.53 to 0.70 ($p$ = $2.2 \times 10^{-4}$, $BF_{10}$ = 160, $n$ = 23, RMSE = 0.014).

These patterns suggest that task-specific response requirements modulate the predictive benefit of early gaze signals. For instance, BB requires a view followed by an interaction (click), and DD primarily involves movement with an optional view. In contrast, NK requires a combination of movement, view, and interaction. The necessity to execute all three response types in NK likely increases the influence of early gaze signals on overall task performance.

To assess whether pd RR adequately captures working-memory demands, we compared its threshold parameter $\boldsymbol{\psi_\theta}$ against the NIHT LSWM threshold parameter $\boldsymbol{\psi_\theta}$. As shown in **Figure 8**, RR scores displayed exceptionally strong positive correlations with both LSWM Form 1 ($r = 0.73$, $p = 7.8\times10^{-4}$, $BF_{10} = 54$, $n = 17$) and LSWM Form 2 ($r = 0.92$, $p = 1.1\times10^{-7}$, $BF_{10} = 1.1\times10^{5}$, $n = 17$) indicating robust convergence between the two approaches. Notably, the latter association (Form 2) was among the highest observed in our study overall, suggesting that the RR task taps core working-memory processes to a degree highly compatible with standardized NIHT working memory assessment.

# Study 2: Test-Retest Reliability

**Figure 9** depicts test-retest plots for the mean RTs (Session 1 vs. Session 2) in NK, DD, and BB, with ICC values of 0.71 ($n = 31$, $p = 2.2\times10^{-6}$), 0.76 ($n = 37$, $p = 3.9\times10^{-11}$), and 0.85 ($n = 34$, $p = 3.5\times10^{-11}$), respectively, all surpassing conventional thresholds for strong reliability. **Figure 10** provides analogous plots for gRTs, revealing comparable ICCs overall. NK showed a negligible increase (0.71 to 0.72), while DD (0.76 to 0.73) and BB (0.85 to 0.82) decreased slightly more. These high reliabilities are particularly notable given that sessions took place in a hospital setting, where external interruptions and participant fatigue are more common than might be expected in a controlled laboratory environment. Within a given task repeat, intersession intervals ranged from minutes to weeks. Median intersession intervals, computed separately for same-task test-retest pairs (NK, DD, BB), ranged from 0.31 to 0.76 days. Longer intervals predicted greater RT change ($r = 0.21$, $n = 51$ participants [99 pairs], $p = 0.039$), though this association was reduced and nonsignificant for intervals under half a day ($r = 0.16$, $n = 30$ participants [48 pairs], $p = 0.27$). Further, while most minigame data collection was supervised, no strict correction of misdirected participant task input was provided, unlike the NIHT tasks. Together, these results imply that process-informed endpoint metrics, such as gRT, can be derived from pD minigames maintain robust trait-like measurement properties even in unpredictable data collection environments.

## Movement Trajectory Analysis

In addition to evaluating task endpoints, we were interested in how much information about individual performance was retained in movement trajectories within the virtual environment. We therefore employed an exploratory UMAP and DBSCAN procedure to perform an unsupervised cluster analysis on Nether Knight movement and viewing trajectories to assess whether distinct behavioral patterns could be identified. As shown in **Figure 11**, four consistent average trajectories emerged, defined by whether participants veered away before selecting their target (indirect vs. direct) and whether they visually fixated on that target early (gaze vs. no-gaze). Two Types (Indirect) exhibited a pronounced lateral deviation, whereas the other two (Direct) followed a more linear path. Within each of those major categories, one subset fixated on the knight before selection (Gaze), while the other maintained a primarily forward-looking orientation and relied heavily on movement cues (No-gaze).

When projected into a 2D latent space via UMAP (see **Figure 12**), each of the four types (Direct+Gaze, DG; Indirect+Gaze, IG; Direct+No gaze, DN; Indirect+No gaze, IN) occupied a distinct region of the embedding, reinforcing the idea that path geometry and visual alignment patterns differ in systematic ways across trials. In total, we found 3527 trials in DG, 834 in IN, 5356 in IG, and 1071 in DN, indicating that roughly two-thirds of Nether Knight trials involved an active gaze fix (DG + IG), and just over half took an indirect route (IG + IN).

We performed a local sensitivity analysis around the reported UMAP/DBSCAN parameters. Across 81 nearby parameter combinations, identifiability ranged 39–62%, the number of clusters ranged 2–7, and cluster-validity indices remained in a consistent range (silhouette −0.016 to 0.28; Davies–Bouldin 0.51 to 1.1.

Although these categories reflect trial-level differences, we also examined whether each participant's distribution of trial types was stable enough to serve as a personal profile. To do so, we encoded each session as the proportion of trials belonging to each Type and computed the Jensen-Shannon distance (JSD) between sessions. To visualize whether each session's distribution could "identify" its participant, we constructed a weighted confusion matrix by ranking each session's profile against every participant's average, which is given in **Figure 13**. If no individual-specific information were contained in the trajectory data, the matrix would be uniformly filled. Diagonal entries of the matrix reached or exceeded 50% for many participants, suggesting that these simple movement and gaze-based clusters provide a moderately distinctive signature at the individual level.

Across sessions, most participants showed consistent mixtures of the four trajectory types, and these mixtures were distinctive enough to identify many participants at rates well above chance ($p < 1\times10^{-5}$) with a large effect size (Cohen's $d = 7.6$). A few individuals with unusual or sparse data showed less stable profiles, but the overall pattern indicates that movement path shapes and avatar-gaze-use behaviors captures participant-specific signatures that remain consistent across testing.

These results from this exploratory analysis demonstrate that how one maneuvers and fixates in a simplified choice environment (such as selecting the correct knight under stimulus conflict) reveals stable, individual-level differences that persist across repeated sessions. They also highlight the potential of utilizing more process-level features (e.g., path geometry, gaze shifts) rather than focusing solely on final RTs, complementing our earlier findings on classical RT endpoints. Similar stable movement and fixation patterns have been reported in a Corsi-style block-tapping paradigm (de Sardenberg Schmid & Hardiess, 2024), reinforcing the idea that individuals exhibit robust behavioral patterns across sessions. Those findings also highlighted distinct implicit vs. explicit visual coding strategies, mirroring our observation of gaze-based approaches (direct-gaze vs. no-gaze) and underscoring how such behaviors can reflect individual differences in underlying processes or strategies. Most importantly, the process information for these tasks was recorded natively within the 3D interactive virtual environment, requiring no special equipment or laboratory setups that would be required to track similar information in the real world.

# Discussion

Our validation findings indicate that gamefully designed tasks in a 3D virtual interactive environment can support group-level inference by yielding endpoints that converge with standardized NIHT measures and show high test-retest reliability. Despite the potential complicating factors in a hospital setting, such as medication effects, visitor interruptions, and varying emotional states, reliability remained very high for pD. Notably, reliability was strongest for retests within a half day; longer intersession intervals introduced modest drift in RT estimates. A key advantage of this overall approach is its capacity to capture naturally occurring within- and between-subject variability without sacrificing measurement quality. By moving beyond the highly controlled, often artificially stable conditions typical of standard tasks, pD provides a more ecologically valid assessment context through sustained engagement and continuous spatial navigation that better approximate the complexity of real-world cognitive demands. While the task stimuli themselves remain abstract or gameful rather than given as everyday objects, this interactive approach moves toward bridging the psychometric rigor of standardized measures and the behavioral richness of naturalistic performance. This advantage was especially evident in our study's context, where real-world challenges did not compromise reliability. The interactive nature of the environment appeared to maintain participant motivation across multiple trials, enabling a sufficiently large trial count for each task and supporting stable assessments of individual differences over time.

We emphasize that the pD minigames were not intended to precisely mirror NIHT tasks but rather to function as parallel test instruments designed to integrate richer, game-like elements. For example, BB introduces additional attention and inhibition components beyond those in the standard FICA task, and DD incorporates complex rule-switching beyond the dimensional shifts in the NIHT DCCS. Design choices prioritized for maintaining engagement, providing richer process data for analysis, and setting the stage for multi-step task designs, did not undermine the system's ability to capture core cognitive abilities. Indeed, the strong correlations with NIHT counterparts affirm that these added task complexities can coexist with sound psychometric properties, revealing how individuals respond to more dynamic cognitive demands in the process

Our findings imply that task-specific response requirements modulate the predictive benefit of early avatar gaze signals. For example, in BB the response sequence involves a view followed by an interaction, while in DD the primary action is movement, with the view being optional. In contrast, NK requires movement, view, and interaction, yet gRT yielded the largest convergence improvement, indicating that early avatar gaze commitment can index decision timing even when multiple response modalities must be coordinated. These results underscore the importance of aligning game mechanics with specific measurement objectives to ensure that process-level data accurately reflect the intended cognitive process.

Moreover, our analysis of continuous movement and gaze trajectories affording both gaze and navigation revealed distinct, individualized behavioral profiles in NK. Some participants quickly aligned their avatars and fixated on the correct target, while others delayed fixation until the final moments. These findings mirror known differences, even in highly reductionist tasks, in visually implicit versus explicit rehearsal strategies (de Sardenberg Schmid & Hardiess, 2024), behaviors opaque to the experimenter in the real world without eye tracking. This finding underscores how process-based approaches offer deeper insight than aggregate endpoint metrics (e.g., mean response times, accuracies, thresholds) alone, and how added trial depth helps delineate stable individual differences in strategy or execution.

A notable advantage of pD implementation over, for instance, the NIHT is the ability to collect behavioral data with minimal or nonexistent supervision. Furthermore, the only hardware needed is available off-the-shelf at consumer prices and requires no special setup. While human gaze in the real world can classically be extracted with eye tracking (Linari et al., 2022), substituting avatar gaze in a 3D virtual interactive environment makes these analyses readily scalable outside the laboratory without special equipment or onerous calibration routines.

Regarding scalability and accessibility, participants in our study varied widely in their gaming backgrounds, from seasoned Minecraft players to complete novices. Because the minigames are self-contained, however, any familiarity with Minecraft's broader sandbox features was functionally irrelevant with respect to performance on the core assessments. Mastery of the simplified minigame mechanics proved straightforward for all participants, with no evident learning effects across sessions (see Figure 10). As a result, scalability of remote testing permits the gathering of large trial counts under less restrictive testing conditions.

The hospital environment was necessary for our underlying study linking brain activity with behavior and included the added benefit of diverse cognitive and educational backgrounds for the study participants. Nonetheless, this is a clinical pediatric population in a clinical setting, elevating potential between- and within-subject differences. Future studies should replicate and extend these findings in more varied settings and populations, including healthy community samples as well as under more controlled laboratory conditions. Refining enrollment protocols and tutorials may help address potential selection effects or confounds while still leveraging the high task engagement we observed.

Several factors limit full generalizability. First, participants were recruited from a pediatric neurology unit rather than a healthy community sample; covariate-adjusted analyses suggest epilepsy and medication status do not confound the primary findings (Supplementary Materials, Section 2), but replication in non-clinical populations is warranted. Second, the 5–24 age range introduces developmental heterogeneity; age-stratified analyses would require larger samples and fall outside the present scope. Third, task completion varied due to unpredictable hospital discharge timing rather than systematic selection by ability, though selection effects cannot be definitively excluded. Fourth, gaming experience showed no significant association with pD performance (Supplementary Materials, Section 3). Finally, longer intersession intervals predicted modest RT drift, an effect concentrated in gaps exceeding one day. Within one day, the test-retest reliability does not detectably differ. Additional details on post-hoc power (Supplementary Materials, Section 4), task completion patterns (Section 5), intersession interval effects (Section 6), per-figure demographics (Section 7), self-selected engagement rates (Section 8), trial-count sensitivity (Section 9), and all-session data quality (Section 10) are provided in the Supplementary Materials.

Ongoing iterations of pD will expand beyond brief minigames to more dynamic, complex tasks. By integrating puzzle-solving, spatial navigation, and evolving objectives, new minigames can capture the complex, adaptive behaviors that single-response paradigms rarely address. Critically, such designs would allow probing of real-time planning, goal-directed actions, and flexible outcome evaluations—cognitive processes often linked to fronto-striatal circuits (Morris et al., 2016). Coupled with the high-dimensional time-series logging of pixelLOG, these more expressive tasks should provide deeper insights into how participants map their actions to plans and utilities, illuminating higher-level cognitive processes that traditional interfaces rarely expose. Crucially for researchers who adopt the pD platform, the same codebase, logging, design features, instructions, themes, gameplay mechanics, contexts and analytical pipelines could be used for both classically reductionist tasks, such as those present here, and novel complex, multi-step, expressive tasks (COMETs).

Taken together, our results demonstrate that simple minigames defined in a 3D virtual interactive environment equipped with comprehensive behavioral logging can reflect well-known cognitive constructs while also revealing individualized differences in task processes. This joint achievement of psychometric rigor and improved ecological richness supports the development of more dynamic, engaging methods of cognitive assessment adaptable to diverse populations and settings. With continued refinement of data-processing pipelines and collaboration with assistive-technology developers, 3D virtual interactive environments such as pD, may help realize the full potential of such testing media in neuropsychological research, particularly for capturing individual differences through increased trial count and depth.

# Conclusions

Despite substantial variability inherent to the hospital environment, pD minigames yield reliable, construct-consistent measures of executive function that converge with established NIHT benchmarks. These convergent results confirm that the core constructs (rule shifting, inhibitory control, and working memory) remain intact within pD's more naturalistic task formats. Moreover, capturing continuous movement and gaze patterns in these richer environments offers new process-level insights that transcend endpoint metrics. By blending comprehensive behavioral logging with an interactive, user-friendly interface, pD provides a template for how future cognitive assessments may be designed and deployed in varied contexts. These results suggest cognitive assessments can maintain the rigor of standardized measures while leveraging continuous behavioral data to capture how individuals plan, decide, and adapt, which is something traditional task formats miss.

# Declarations

## Funding

This study was supported in part by the Washington University OVCR Seed Grant and Here and Next Programs, the St. Louis Children’s Hospital Foundation, NIH T32NS115672 and NSF NAIRR240235.

## Conflicts of Interest/Competing Interests

The authors declare that they have no conflicts of interest or competing interests.

## Ethics Approval

All procedures performed in this study were approved by the Institutional Review Board at Washington University in St. Louis (IRB # 202405134) and adhered to the ethical standards of the 1964 Declaration of Helsinki and its later amendments or comparable ethical standards.

## Consent to Participate

Written informed consent was obtained from all participants (or their parents/guardians in the case of minors) prior to their inclusion in the study.

## Consent for Publication

Not applicable.

## Availability of Data and Materials

Project information, including the data necessary to replicate these analyses and how to access the games themselves, can be found at https://osf.io/r6jcn/.

## Code Availability

The code necessary to replicate these analyses can be referenced at https://osf.io/r6jcn/.

## Authors' Contributions

**Dom CP Marticorena:** Conception and design of the study, data analysis, manuscript drafting.

**Zeyu Lu:** Software development (pixelLOG), data collection, manuscript editing.

**Chris Wissmann:** Data interpretation, critical manuscript review.

**Yash Agarwal, David Garrison, John Zempel:** Participant recruitment, clinical data coordination, critical review of manuscript.

**Dennis L. Barbour:** Overall supervision, secured funding, manuscript finalization.

All authors read and approved the final manuscript.

### Open Practices

The authors followed guidelines recommended by the American Psychological Association (APA) for reporting methods and analyses.

## Acknowledgments

We thank Carrie Cao and Alex Magrath for assistance with data collection, as well as staff at the Washington University Pediatric Epilepsy Center.

# Figures and Tables

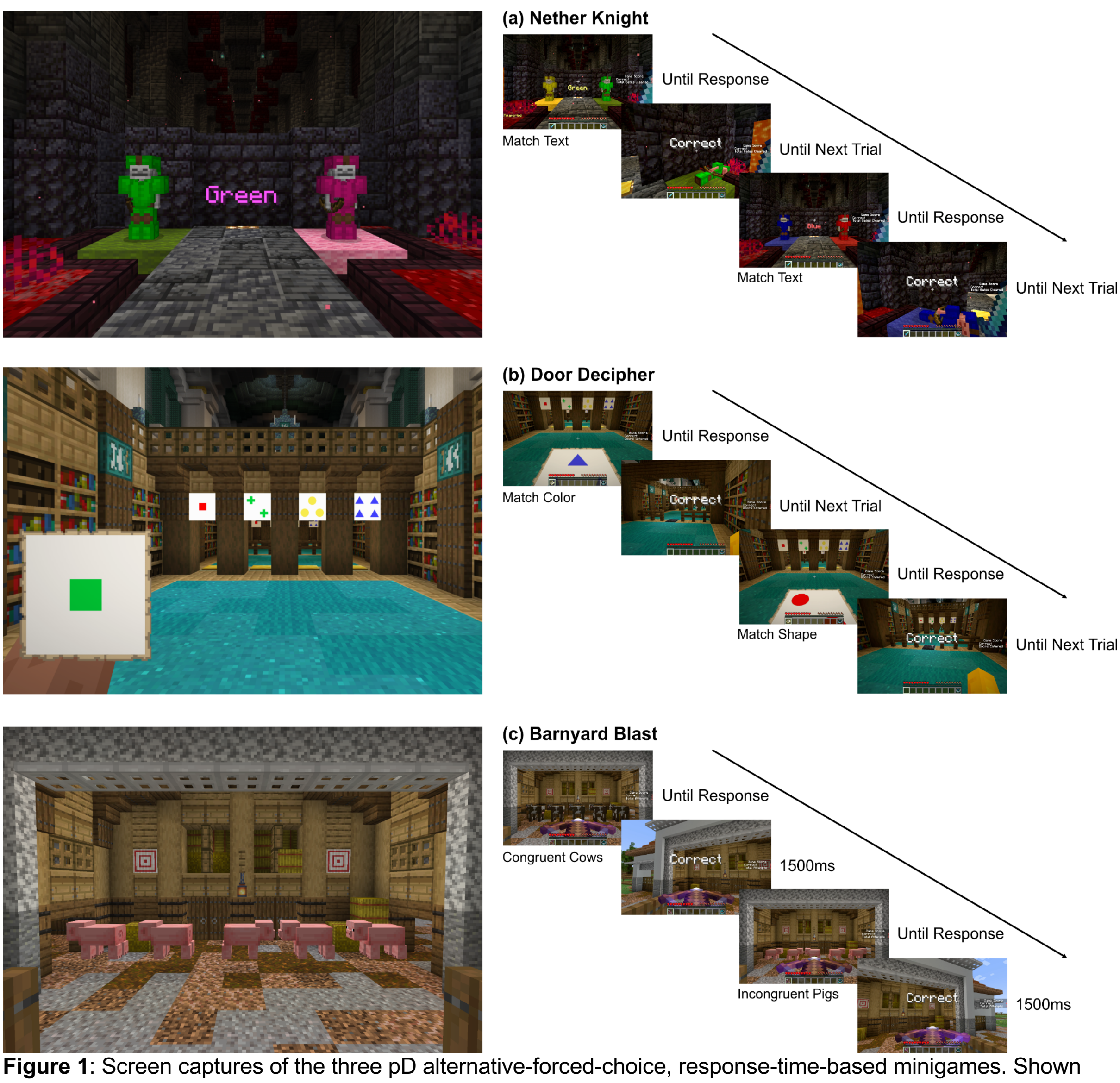

**Figure 1**: Screen captures of the three pD alternative-forced-choice, response-time-based minigames. Shown from top to bottom: Nether Knight (NK), which requires participants to respond to an incongruent colored word stimulus by interacting with a matching knight; Door Decipher (DD), in which participants classify key cards by color, shape, or quantity and enter the corresponding door; and Barnyard Blast (BB), which tasks participants with tagging the left or right target based on animal orientation. Each minigame includes feedback and a brief inter-trial interval before the next trial begins.

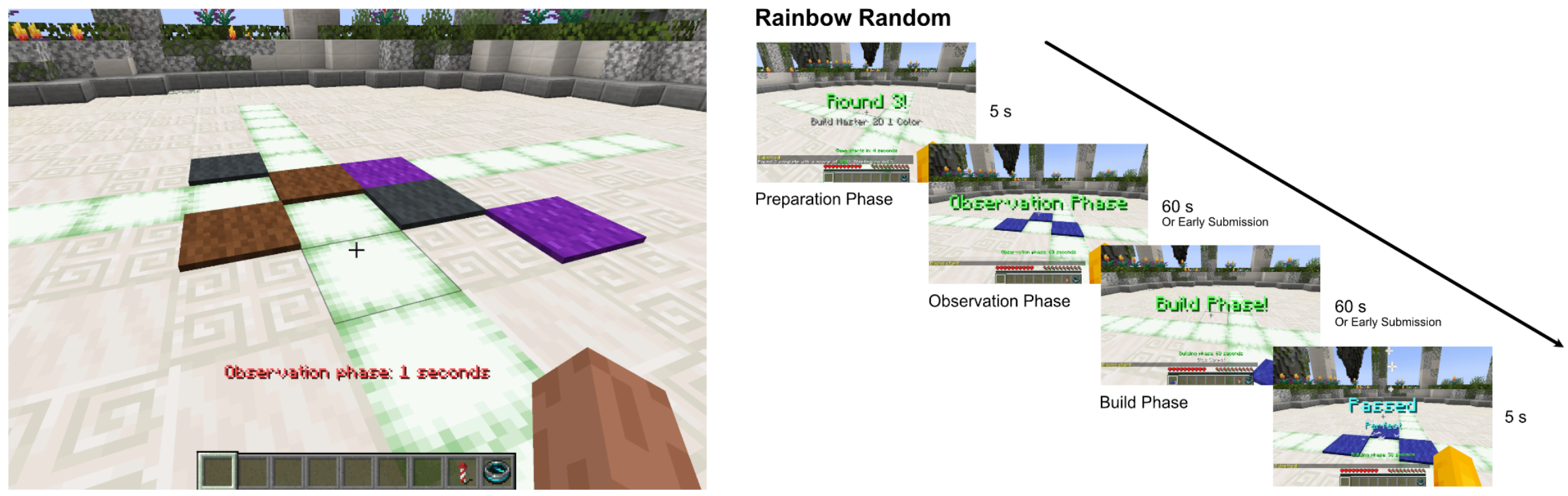

**Figure 2**: Screenshot illustrating Rainbow Random (RR), an adaptive working memory task in which participants reconstruct a 2D pattern composed of up to three colors. Each trial includes a brief preparation phase, a 60-second observation period (which can be shortened by participant), a 60-second build phase (which also can be shortened) to place colored blocks in the correct arrangement, and a judging phase with pass/fail feedback. Difficulty increases or decreases automatically based on the participant's performance, allowing for personalized assessment of visuospatial working memory capacity.

**(a) Navigation Segment**

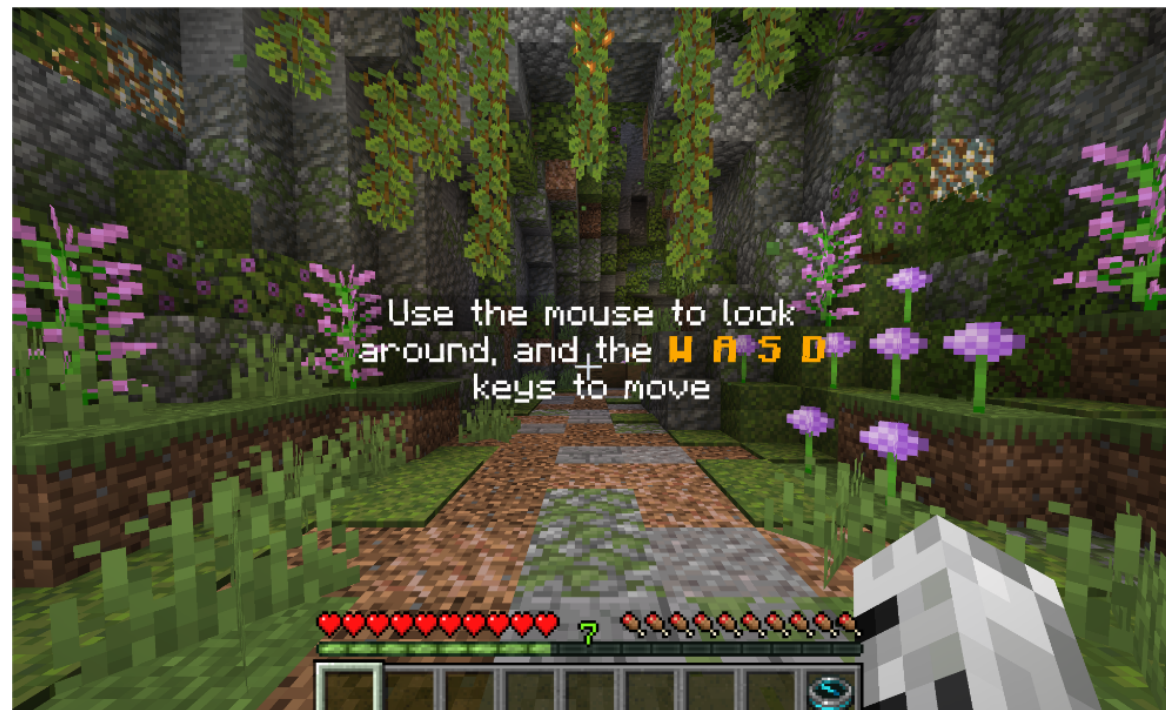

**(b) Target Selection Segment**

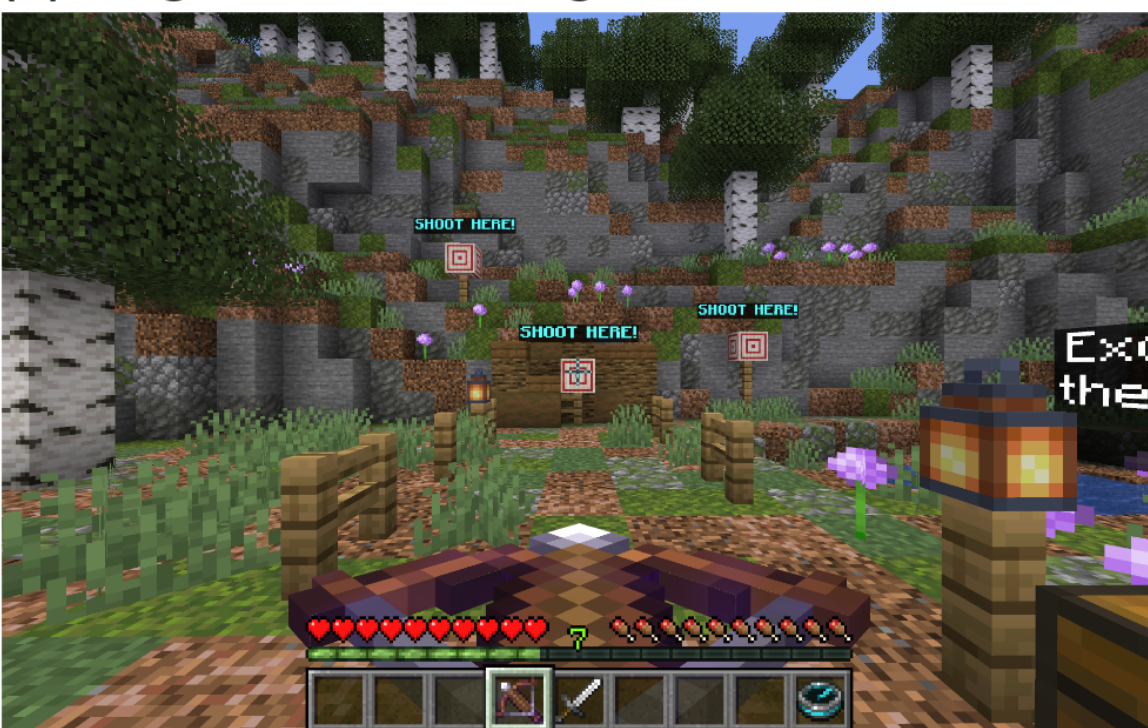

**(c) Block Place Segment**

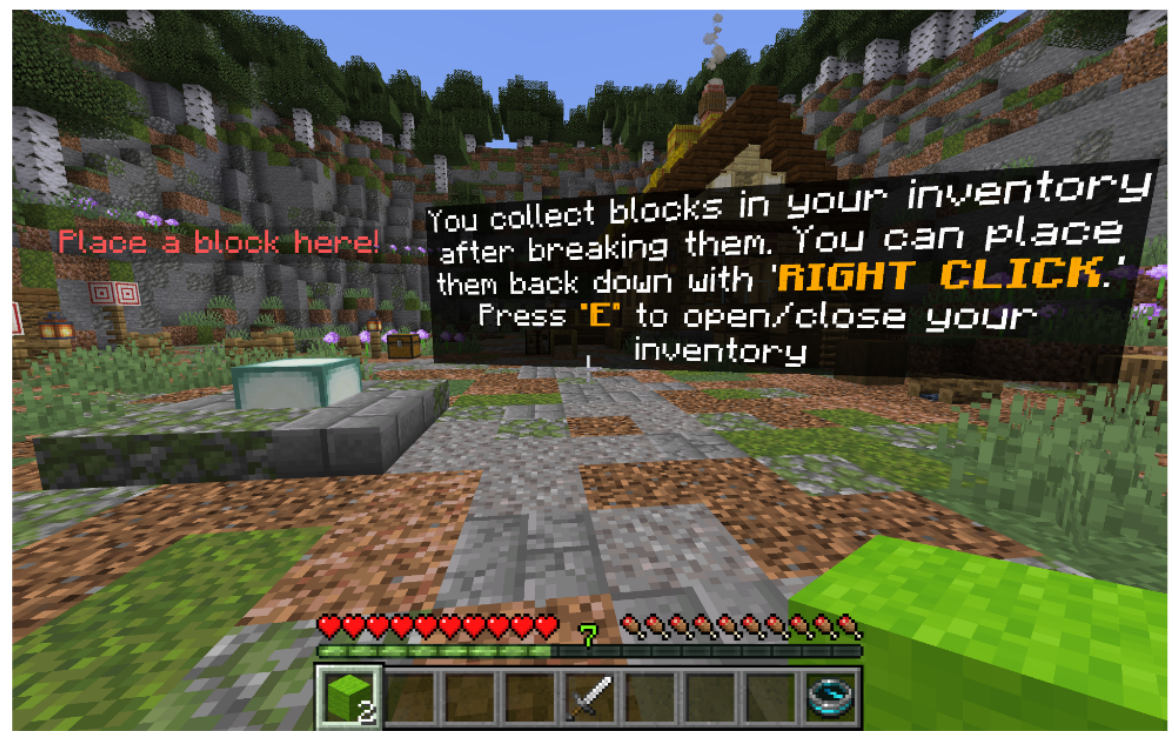

**(d) Block Break Segment**

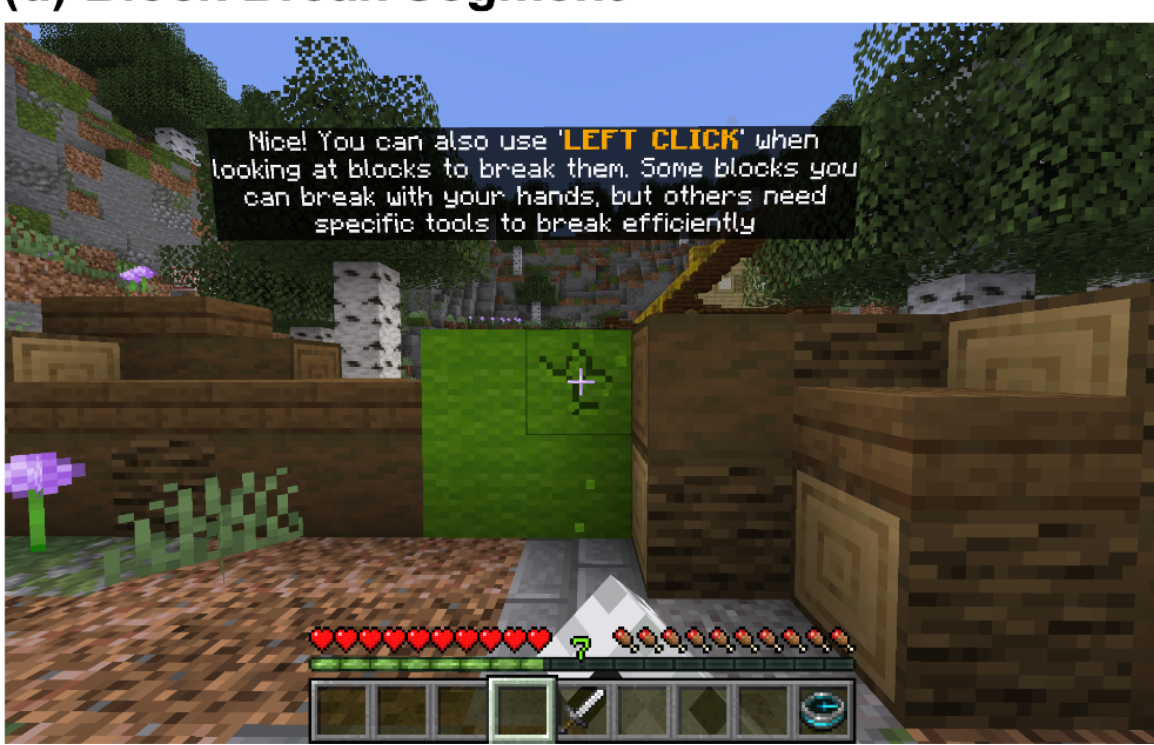

**Figure 3:** Screenshots from representative pD tutorial segments. (a) Navigation Segment: Participants learn basic movement (forward, backward, left, right, looking around) and practice jumping/sprinting. (b) Target Selection Segment: Participants select and use an item from their quick-access bar to interact with an in-game block. (c) Block Place Segment: Participants learn to place blocks retrieved during the tutorial. (d) Block Break Segment: Participants practice breaking blocks with appropriate tools and collect dropped items.

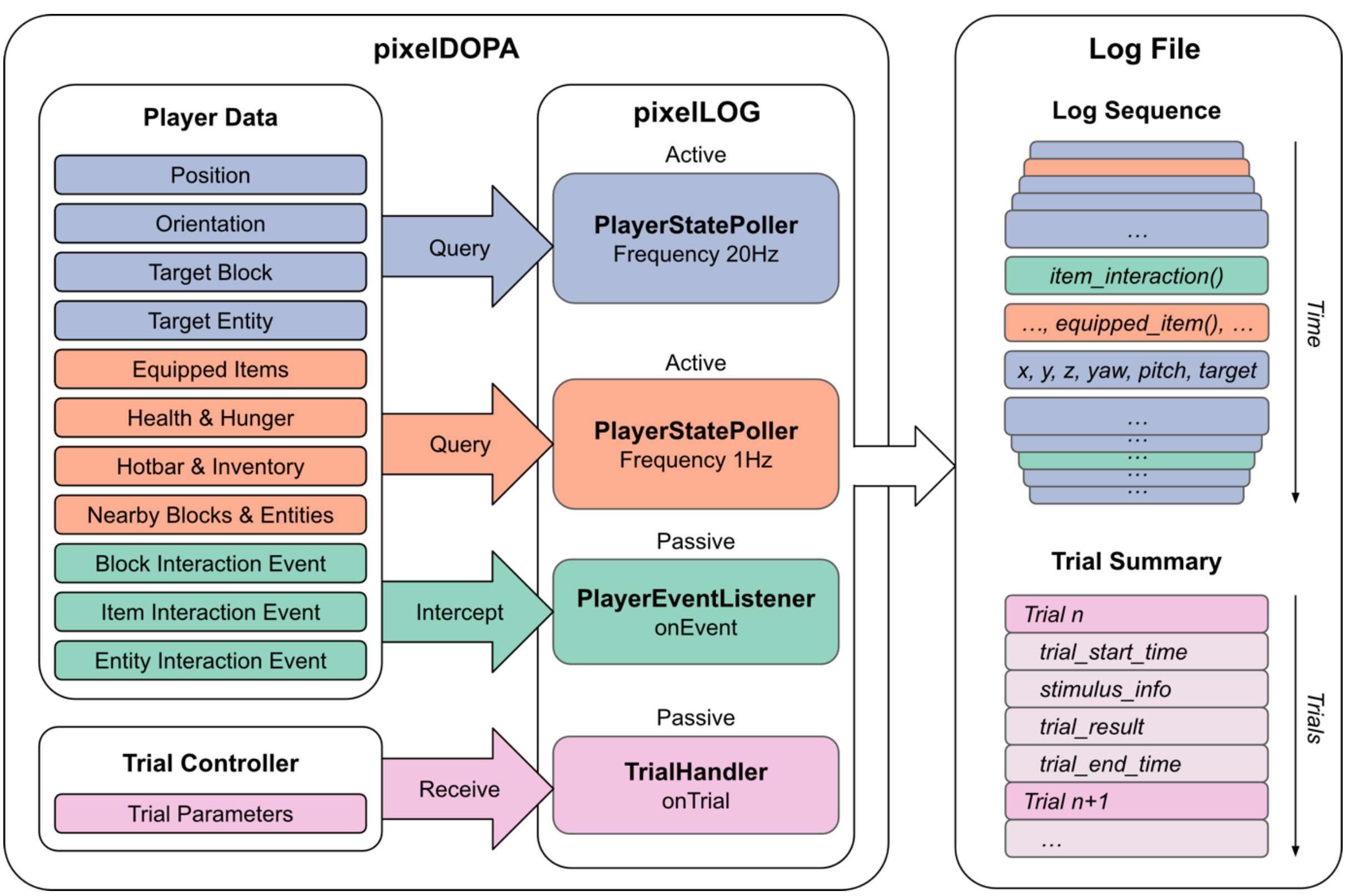

**Figure 4**: pixelLOG coordinates three primary mechanisms: 1) two PlayerStatePollers (operating at 20 Hz and 1 Hz) to continuously sample player states from the pD environment, 2) a PlayerEventListener to intercept discrete, event-based interactions, and 3) a TrialHandler to receive Trial Parameters from the Trial Controller, which is a pD component responsible for managing trial lifecycles and progression. Three types of player data are organized into a time-ordered Log Sequence, while Trial Parameters are structured into a Trial Summary. Together, these components constitute a complete LogFile designed for real-time or offline analysis.

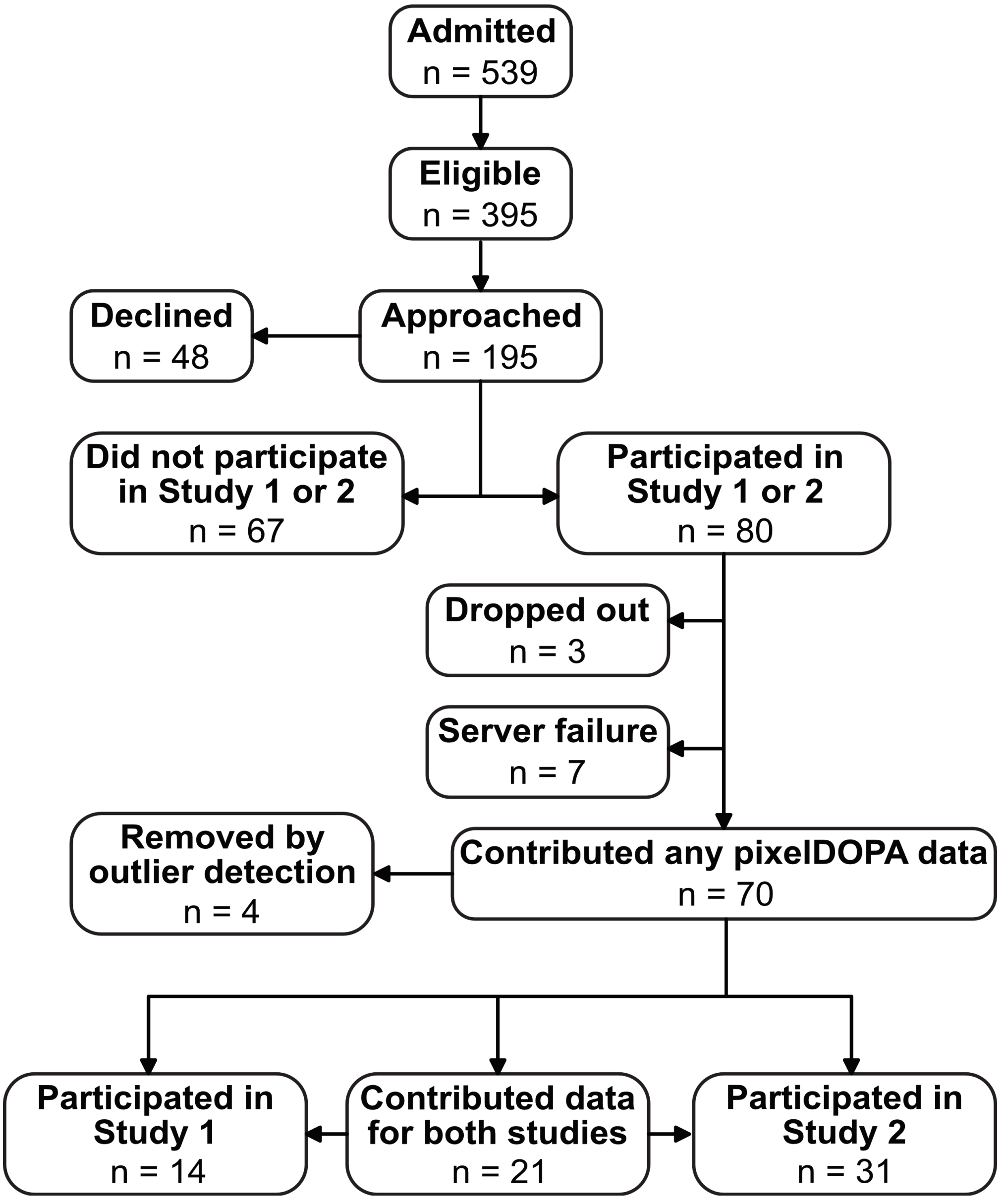

**Figure 5:** Flow diagram illustrating recruitment, enrollment, and final data usage over the fourteen-month study period. Of 539 individuals admitted to the **EEG** Monitoring Unit, 395 were eligible and 195 were approached; 48 declined and 147 consented. Of the 147 consented, 80 consented in Study 1 or Study 2. Data from the first 7 enrollees were unprocessable due to server errors, and 3 dropped out following consent, leaving 70 participants (ages 5–24) enrolled. Of these, 14 participated in only Study 1 for convergent validity, 31 participated in only Study 2 for test-retest reliability, and 21 participated in both, leaving 35 individuals in Study 1 and 52 individuals in Study 2, including overlap.

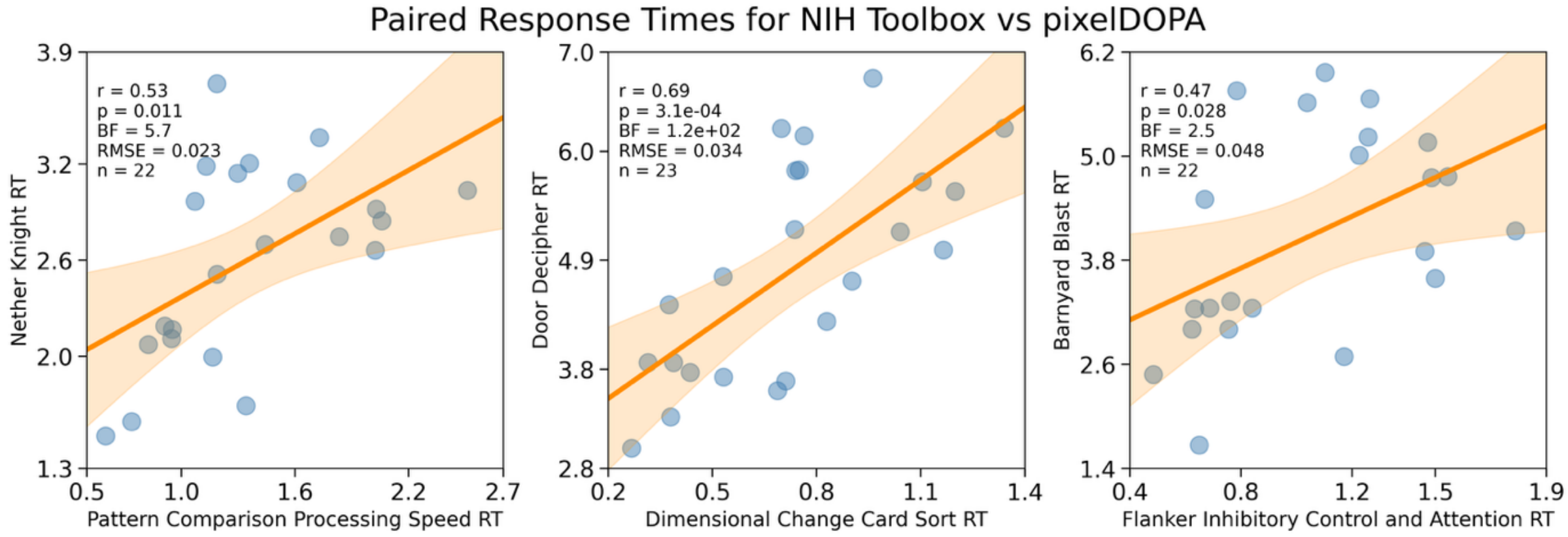

**Figure 6**: Correlations between the mean response times (RTs) recorded for each NIHT cognitive task (horizontal axes) and those recorded for the corresponding pD minigame (vertical axes). Each data point in blue represents an individual participant, with best-fit regression lines (orange) and 95% confidence intervals (orange shading). The Pearson correlation coefficient (*r*), *p*-value, Bayes Factor (BF), root-mean-square-error (RMSE) and sample size (*n*) are reported in each panel. From left to right: 1) PCPS and NK; 2) DCCS and DD; and 3) FICA and BB.

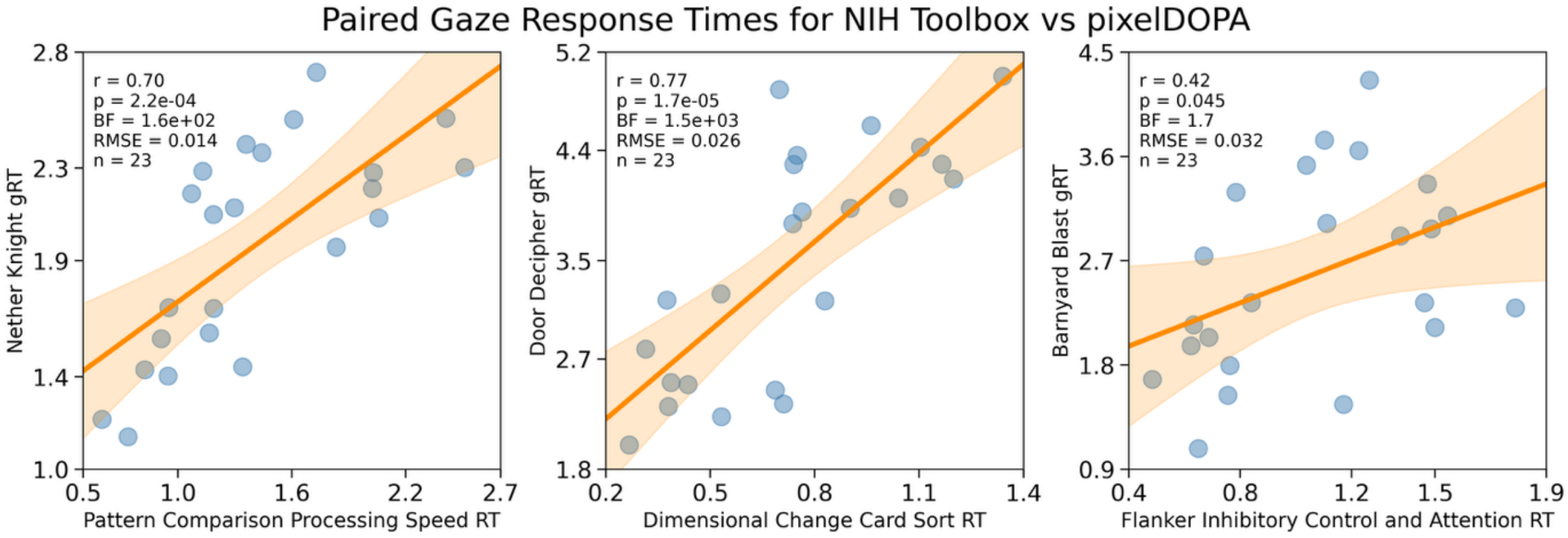

**Figure 7**: Correlations between the mean response times (RTs) recorded for each NIHT cognitive task (horizontal axes) and the corresponding pD minigame mean gaze response times (gRTs) (vertical axes). Plotting conventions as in Figure 6.

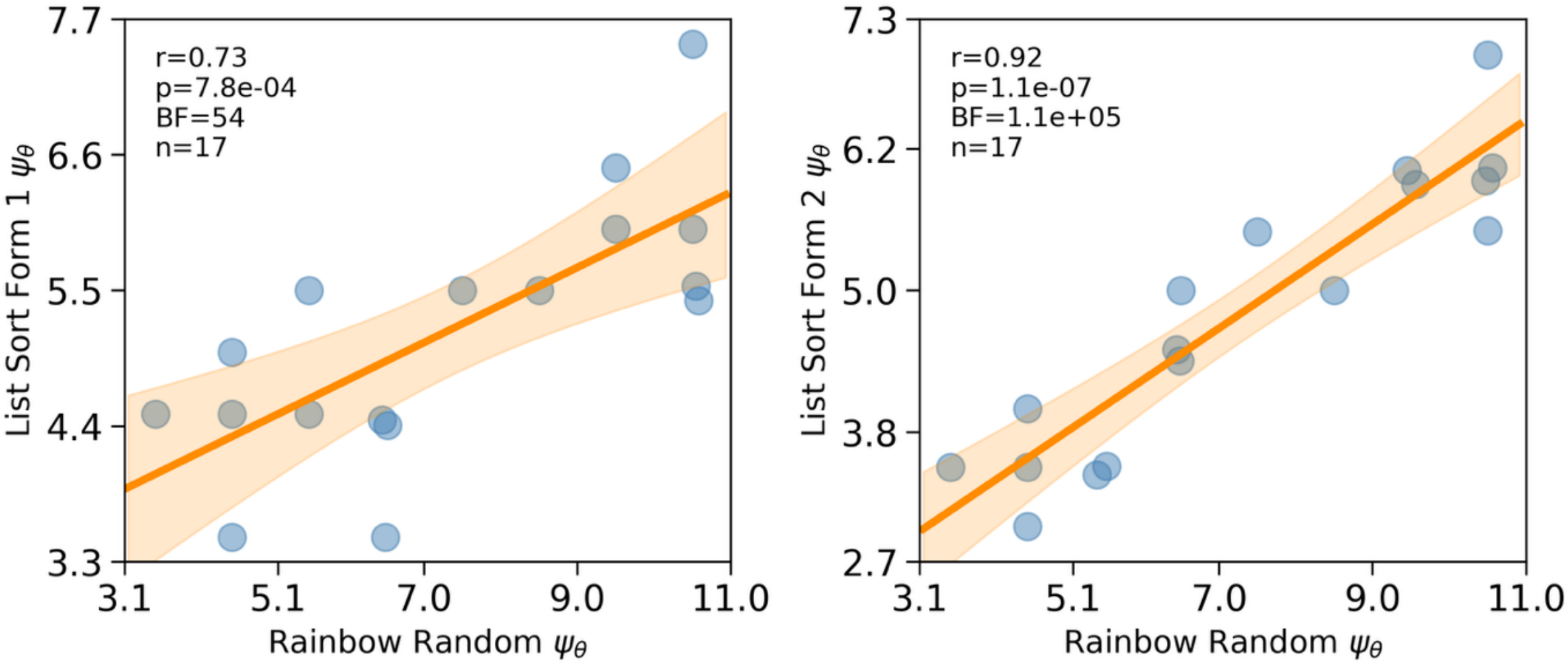

**Figure 8**: Correlations between the psychometric threshold $\psi_\theta$ calculated for NIHT LSWM (horizontal axes) and pD RR (vertical axes). Each data point in blue represents an individual participant with best-fit regression lines (orange) and 95% confidence intervals (orange shading). The Pearson correlation coefficient (*r*), *p*-value and sample size (*n*) are reported in each panel. On the left is the correlation with LSWM Form 1, which requires basic recall, and on the right is correlation with LSWM Form 2, which requires sorting the recalled list according to size. Ceiling effects on the RR tasks may have artificially lowered the correlations.

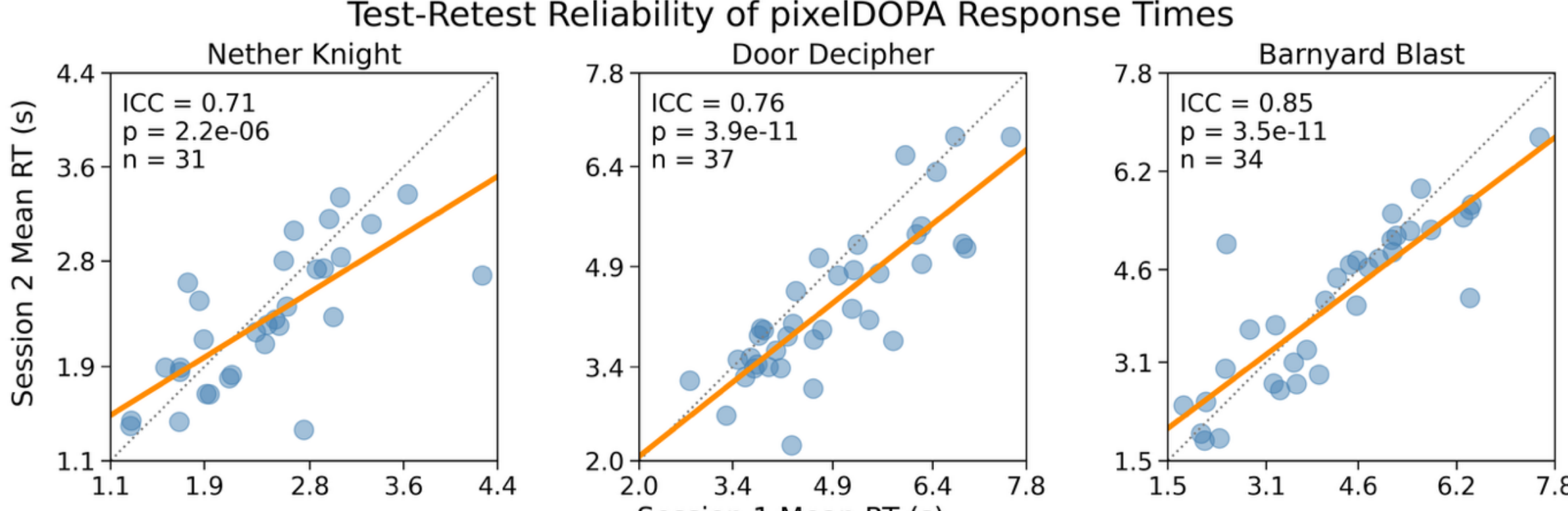

**Figure 9**: Test-retest reliability of the pD minigames, showing Session 1 mean RTs (horizontal axis) versus Session 2 mean RTs (vertical axis) for NK, DD, and BB. Each data point in blue represents an individual participant, with best-fit regression lines (orange). A dotted gray line marks the line of equality ($y = x$). Intraclass Correlation Coefficient (ICC), $p$-value and sample size ($n$) are provided in each panel, illustrating the consistency of RTs across sessions for each minigame.

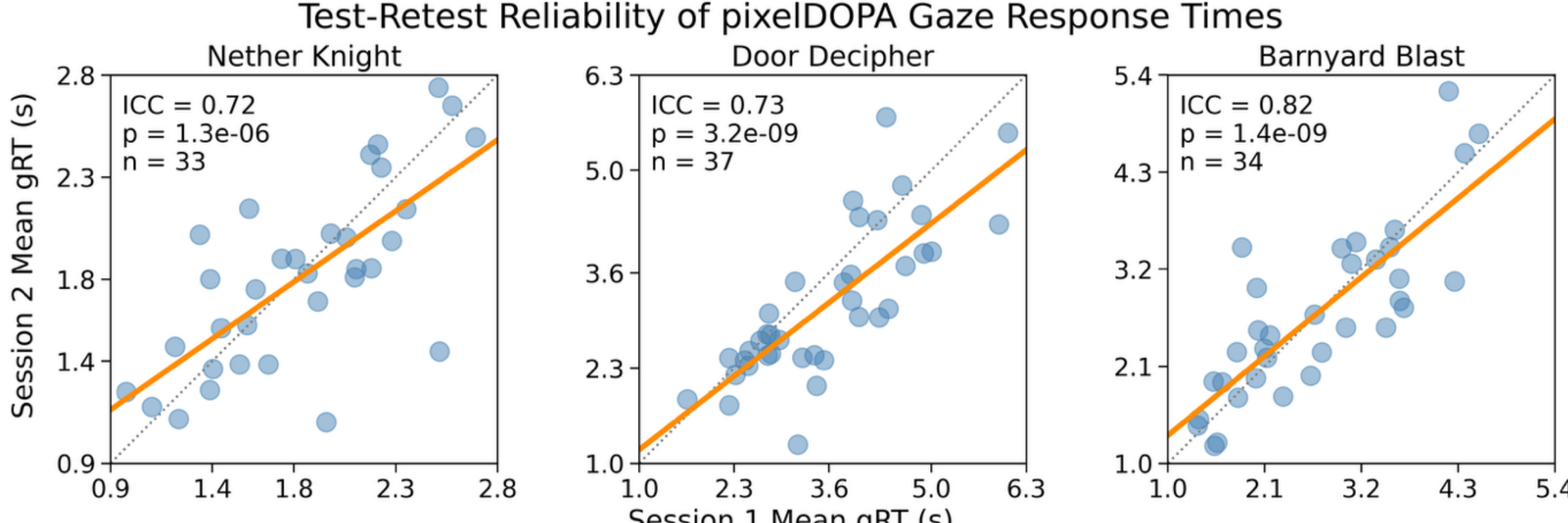

**Figure 10**: Test-retest reliability of the pD minigames, showing Session 1 mean gRTs (horizontal axis) versus Session 2 mean gRTs (vertical axis) for NK, DD, and BB. Plotting conventions as in Figure 9.

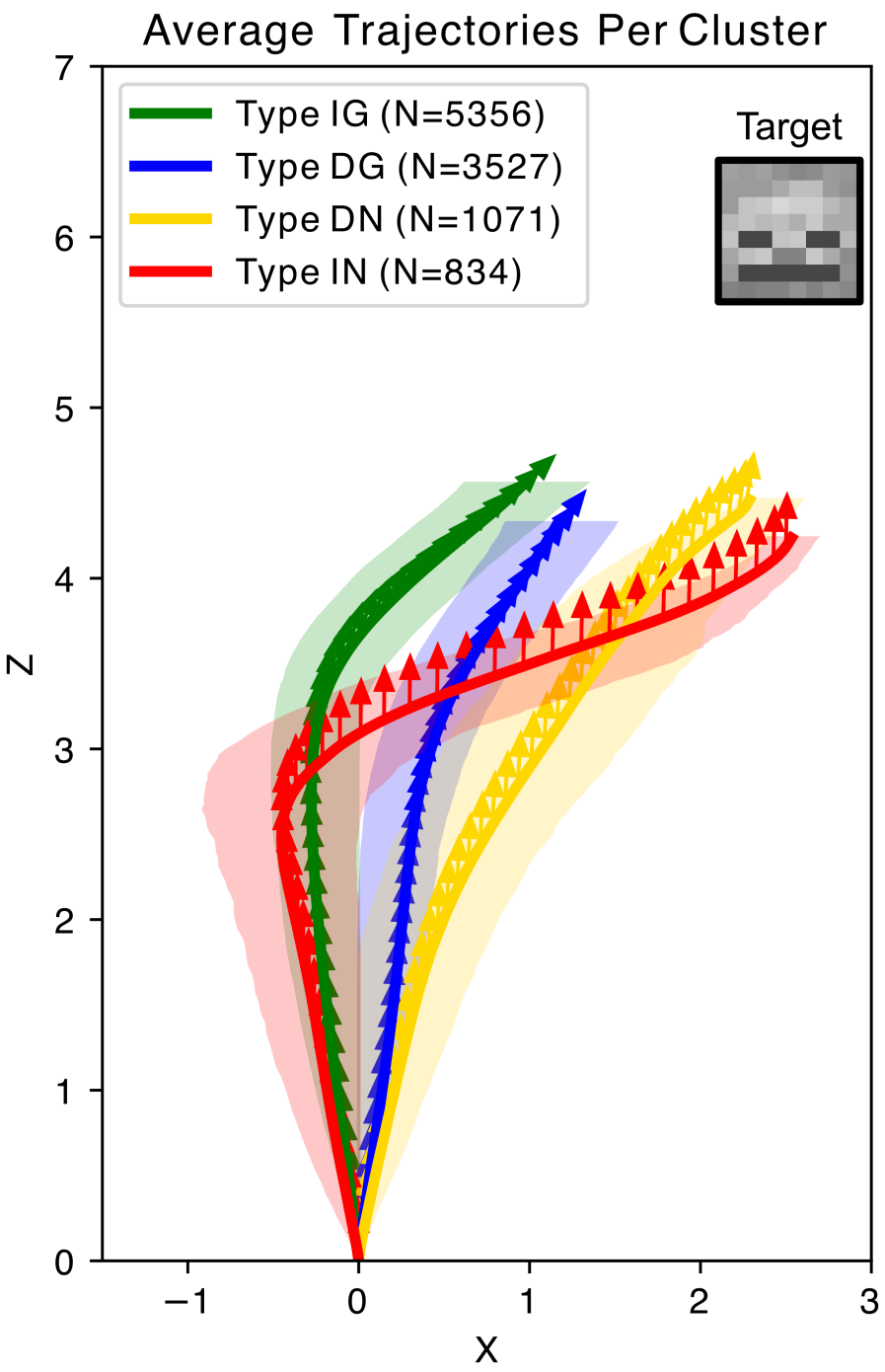

**Figure 11.** Each curve shows the time-aligned, average participant trajectory for one of the four DBSCAN-identified clusters (color-coded), derived from high-dimensional (27D) motion features from all NK trials projected via UMAP. The solid lines indicate mean position in the *x–z* plane at each time step, with shaded regions denoting the 25th–75th percentile range. Arrows, placed at regular intervals, represent the mean heading direction (yaw) of each cluster. The center of mass of the target is shown as a knight face. Cluster membership and count are shown in the legend. This figure demonstrates that each cluster exhibits distinct spatial patterns and heading profiles throughout the duration of the trial.

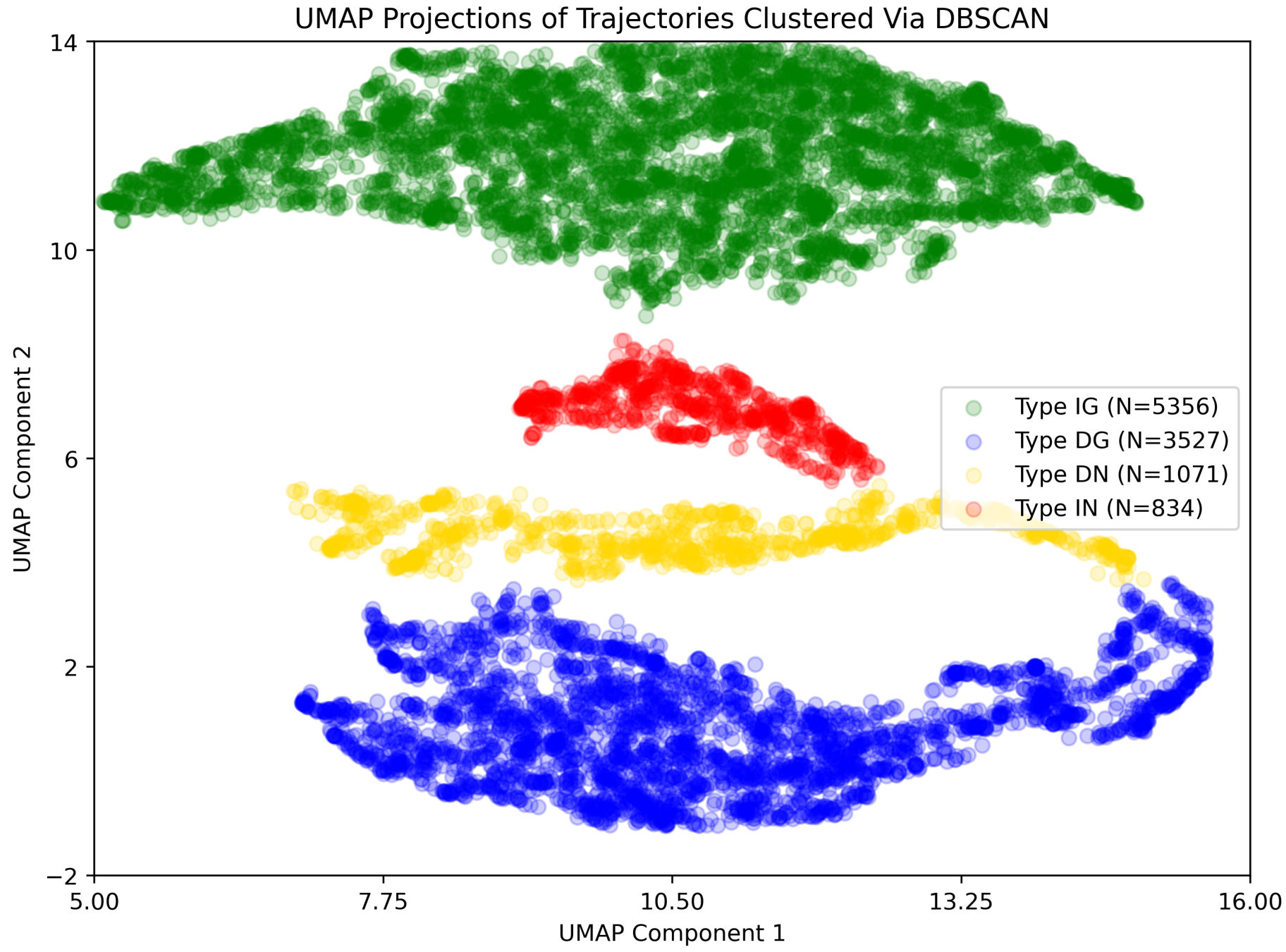

**Figure 12:** The two-dimensional UMAP embedding of the high-dimensional (27D) participant trajectory features. Each point represents an individual trajectory projected into UMAP space, with colors indicating the DBSCAN-assigned cluster. The horizontal and vertical axes represent UMAP Components 1 and 2, respectively. The clustering structure evident in this plot reveals the intrinsic grouping of movement behaviors and suggests that the reduced representation may capture meaningful differences among participants.

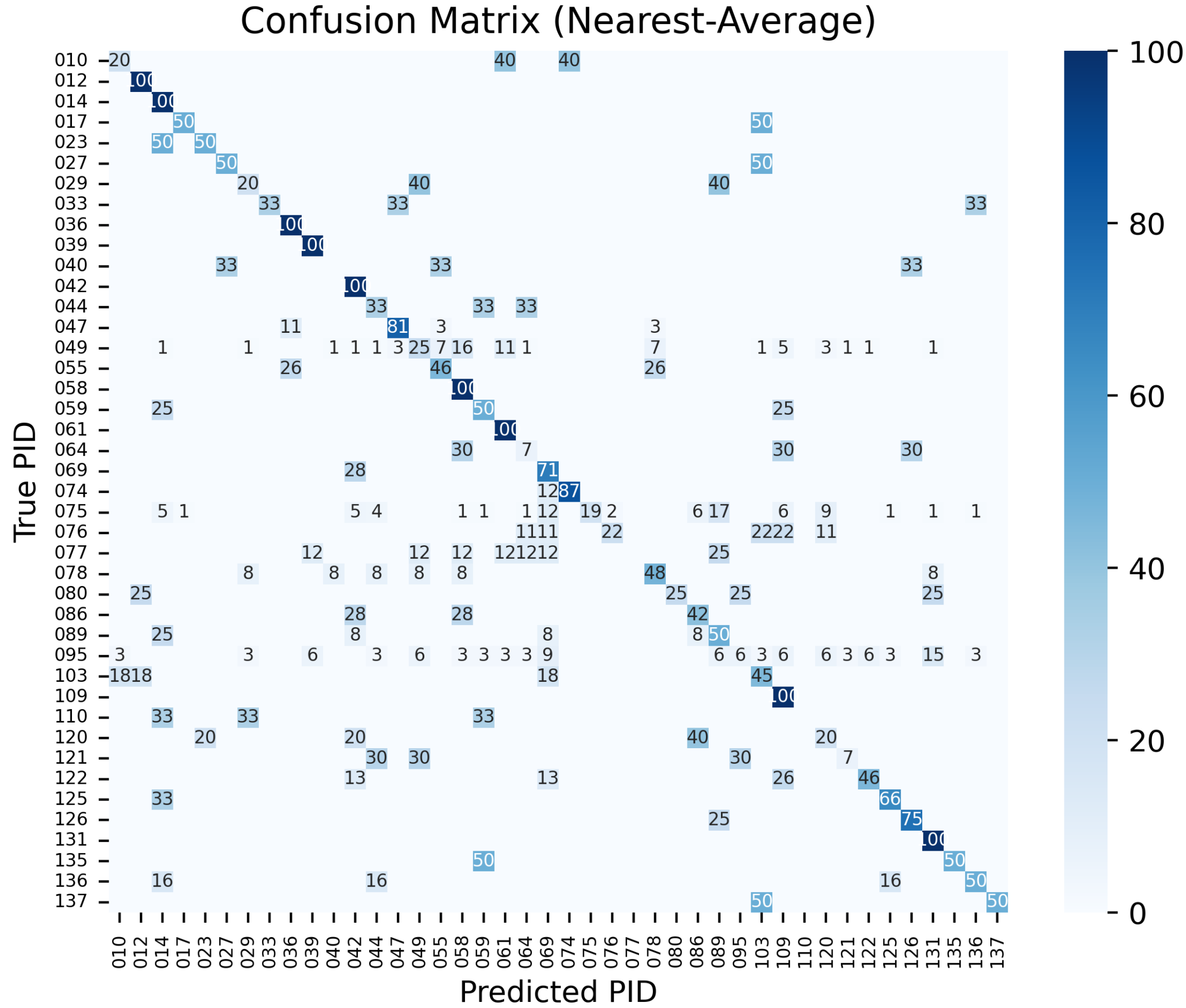

**Figure 13:** The weighted confusion matrix quantifies the identifiability of individual participants based on their Nether Knight movement profiles. For each session, a full vote (1.0) is assigned to the top-ranked (nearest-average) match, with additional partial credits (0.5 for rank 2; 0.25 for rank 3) awarded if the true participant appears further down the ranking. Votes are aggregated for each true participant (row) and normalized so that the sum of votes per row equals 100%. Higher diagonal values indicate strong, distinctive movement patterns leading to correct participant identification, whereas substantial off-diagonal entries point to systematic misclassification or behavioral overlap.

**Table 1**: Overview of pD minigames and their NIHT counterparts. Game mechanics indicate the response modalities required (movement, view adjustment, click selection). Unique elements describe features distinguishing each minigame from its standardized counterpart.

| pD Minigame | Game Mechanics | Cognitive Construct | NIHT Counterpart | Unique Elements |
|---|---|---|---|---|
| Nether Knight (NK) | Movement, view adjustment, click selection | Processing speed, selective attention | Pattern Comparison Processing Speed (PCPS) | Stroop-like color-word conflict; avatar navigation to target |
| Door Decipher (DD) | Movement, view adjustment, no click | Rule shifting, cognitive flexibility | Dimensional Change Card Sort (DCCS) | Rule switches without explicit cue; 3 dimensions (color, shape, quantity) |
| Barnyard Blast (BB) | View adjustment, click selection, no movement | Inhibitory control, selective attention | Flanker Inhibitory Control and Attention (FICA) | Reverse trials (pig = opposite response); congruent/incongruent flankers |
| Rainbow Random (RR) | Movement, block placement | Visuospatial working memory | List Sorting Working Memory (LSWM) | Adaptive staircase; 2D spatial reconstruction; self-paced observation |

**Table 2**: Trial-level data quality for Study 1 (convergent validity). Each row reports the first valid session per participant. Exclusion percentages reflect trials removed under classical RT or gaze-based RT (gRT) criteria. Trials Salvaged indicates net trials gained (+) or lost (−) by switching from RT to gRT criteria. gRT criteria also recovered sessions excluded under RT (2 NK, 2 BB), occasionally adding participants to gRT-based analyses (see Figures 7 vs. 6).

| Task | Eligible | Included in RT (gRT) | Trials | RT Excl. % | gRT Excl. % | Trials Salvaged |
|---|---|---|---|---|---|---|
| PCPS | 25 | 22 | 1415 | 2.0% | — | — |
| DCCS | 26 | 23 | 920 | 1.2% | — | — |
| FICA | 25 | 22 | 660 | 1.7% | — | — |
| NK | 25 | 22 (23) | 907 | 5.3% | 6.6% | –14 |
| DD | 26 | 23 (23) | 997 | 7.6% | 5.5% | +23 |
| BB | 25 | 22 (23) | 988 | 8.7% | 7.1% | +24 |

**Table 3:** Trial-level data quality for Study 2 (test-retest reliability). Each included participant contributes one test-retest session pair. Included (RT) and Included (gRT) differ for NK because gRT recovered 4 sessions excluded under RT, adding 2 participants to gRT-based reliability analyses (see Figures 10 vs. 9). Negative Trials Salvaged indicates gRT criteria were stricter at the trial level for that task.

| Task | Eligible | Included in RT (gRT) | Trials | RT Excl. % | gRT Excl. % | Trials Salvaged |
|---|---|---|---|---|---|---|
| NK | 36 | 31 (33) | 2461 | 2.9% | 4.3% | –38 |
| DD | 37 | 37 (37) | 2451 | 0% | 0% | 0 |
| BB | 34 | 34 (34) | 2430 | 0% | 2.0% | –49 |

# Supplemental Material

## S1. Participant Demographics

The full enrollment pool for this study comprised 80 individuals admitted to the EEG Monitoring Unit over the study period of 14 months. Server failures left 7 participants unable to be processed, and 3 participants dropped out after consenting. The remaining 70 define the analysis cohort used throughout. Four participants had data excluded due to extreme outlier detection.

Study 1 gathered a convergence-only cohort (14) comprised of participants who completed at least one NIHT task and its corresponding pD minigame. Study 2 gathered a reliability-only (31) cohort comprised of participants who completed two or more sessions of at least one pD minigame. A cohort of 21 participants completed both Study 1 and Study 2.

Across the analysis cohort (*n* = 66), participants ranged in age from 5 to 24 years (M = 13, SD = 4.1), with 53% female. Of participants in the analysis cohort, 86% (57/66) had a previous diagnosis of epilepsy. Among those with epilepsy, 74% (42/57) were medicated during their hospital stay. Epilepsy subtypes included unspecified (40%), intractable (28%), generalized (16%), and other focal or mixed presentations (16%). The most common comorbid diagnosis was ADHD (*n* = 13), followed by migraine (*n* = 4).

Minecraft experience was assessed via self-reported familiarity on a 1–10 scale and hours played per week from 0–10+. Only two participants selected 10+ categories, and these responses were coded as 10. Among participants with available data in the analysis cohort (*n* = 42), mean familiarity was 8 (SD = 2.5, range 2–10), and mean hours per week was 2.1 (SD = 3.1, range 0–10).

Hospital stays ranged from 0 to 34 nights (median = 2, mean = 3.3). A total of 15 participants (23%) experienced a seizure during EEG monitoring, but no peri-seizure behavioral data are included in this study.

## S2. Effects of Epilepsy and Medication Status

To evaluate whether clinical characteristics confounded the primary findings, we computed covariate-adjusted correlations after removing variance associated with epilepsy status (yes/no) and medication status (on/off).

### Study 1: pD–NIHT convergence

In the convergence cohort ($n$ = 35), 30 participants had a previous diagnosis of epilepsy and 5 did not. Of those with epilepsy, 19 were medicated during their stay. Adjusted correlations were not meaningfully affected: NK–PCPS $r$ = 0.62 ($p$ = 0.0016, $n$ = 23); DD–DCCS $r$ = 0.62 ($p$ = 0.0013, $n$ = 23); BB–FICA $r$ = 0.52 ($p$ = 0.011, $n$ = 23); RR–LSWM $r$ = 0.94 ($p = 3.1\times10^{-8}$, $n$ = 17).

### Study 2: pD–pD Reliability

In the reliability cohort ($n$ = 52), 47 participants had a previous diagnosis of epilepsy and 5 did not. Of those with epilepsy, 35 were medicated. Covariate-adjusted test-retest correlations between Session 1 and Session 2 outcomes remained strong NK $r$ = 0.72 ($p = 5.6\times10^{-6}$, $n$ = 31); DD $r$ = 0.83 ($p = 1.8\times10^{-10}$, $n$ = 37); BB $r$ = 0.88 ($p = 1.0\times10^{-11}$, $n$ = 34).

# S3. Effects of Minecraft Experience

A primary concern was that Minecraft experience could differentially influence pD performance compared to NIHT performance, confounding convergent validity.

### Experience and pD Performance

Across participants with experience data ($n$ = 28–39), overall pD performance showed no significant association with Minecraft familiarity or hours of gameplay per week. For mean

RT: $r = –0.40$ to $–0.25$ ($p \geq 0.2$). For mean gRT: $r = –0.35$ to $0.0045$ ($p \geq 0.98$). For working memory threshold: $r = 0.078$ to $0.35$ ($p \geq 0.84$).

### Study 1: Experience-Adjusted pD–NIHT convergence

The convergence cohort with available experience data was limited ($n = 11–13$), but adjusted correlations remained directionally consistent: RT $r = 0.47$ to $0.79$; gRT $r = 0.54$ to $0.57$ ($p = 0.0021$ to $0.15$); working memory threshold $r = 0.83$ to $0.95$ ($p \leq 0.0053$). The data collected therefore do not support the hypothesis that Minecraft experience contributes significantly toward explaining pD–NIHT convergence.

### Study 2: Experience-Adjusted Test-Retest Reliability

Covariate-adjusted correlations between Session 1 and Session 2 outcomes, after residualizing on self-reported familiarity and hours played per week, remained strong: RT $r = 0.60$ to $0.82$; gRT $r = 0.64$ to $0.78$ ($n = 21–23$, $p \leq 2.5\times10^{-3}$).

## S4. Power Analysis

Given the observed pD–NIHT effect sizes ($r = 0.42$ to $0.92$, $n = 17$ to $23$), post-hoc power ranged from 0.61 to 0.98 (alpha = 0.05, two-tailed). Bayes Factors provided moderate to strong convergent evidence: $BF_{10} = 1.7$ to $1.2\times10^{2}$ for RT-based correlations; $BF_{10} = 54$ to $1.1\times10^{5}$ for working memory threshold correlations. Detecting medium effects ($r \approx 0.30$) with 80% power would require approximately 84 participants per task pair.

## S5. Task Completion and Selection Effects

Task completion variability resulted primarily from inpatient logistics. In Study 1, 35 participants contributed to at least one of the four task pairs, and 6 completed all four (17%). Completers had longer evaluation time (median of 1 vs 0.28 days, $p = 0.050$) than partial completers. In the analysis cohort ($n = 66$), DD showed the highest dropout rate (17%) compared with the lowest dropout rate, NK (3%). Ordering across pD tasks showed DD occurred last most frequently. Specifically, it came after NK in 67% ($n = 55$), after BB in 62% ($n = 47$), and after RR in 59% ($n = 44$), and was 3rd in median delivery order, while NK was 1st. Furthermore, prior Minecraft experience did not predict completion: neither familiarity ($r = -0.089$, $p = 0.72$) nor hours played ($r = 0.085$, $p = 0.73$) correlated with missingness.

## S6. Test-Retest Interval Effects

Intersession intervals ranged from 0.0021 to 15 days. Median intersession intervals were computed separately for each same-task test-retest pairing (NK, DD, BB) and ranged from 0.31 to 0.76 days. Longer intervals predicted larger RT change ($r = 0.21$, $n = 51$ participants [99 pairs], $p = 0.039$), but this association weakened for shorter intervals: $<1$ day, $r = 0.2$ ($n = 44$ participants [77 pairs], $p = 0.089$); $<0.5$ days, $r = 0.16$ ($n = 30$ participants [48 pairs], $p = 0.27$).

## S7. Per-Figure Demographics

Table S1 summarizes per-panel demographics and trial counts for Figures 6–10, including age range/mean, sex, trial counts, epilepsy status, and medication status.

| Figure | Panel | N | Age mean | Age SD | pD trials mean | NIH trials mean | Female | Male | Epilepsy | Medicated |
|---|---|---|---|---|---|---|---|---|---|---|
| **Fig 6** | DD–DCCS | 23 | 14 | 4.4 | 38 | 40 | 14 | 10 | 19 | 12 |
| **Fig 6** | BB–FICA | 22 | 14 | 4.7 | 41 | 30 | 15 | 8 | 20 | 13 |
| **Fig 6** | NK– PCPS | 22 | 14 | 4.1 | 37 | 61 | 14 | 9 | 19 | 12 |
| **Fig 7** | DD–DCCS | 23 | 14 | 4.4 | 38 | 40 | 14 | 10 | 19 | 12 |
| **Fig 7** | BB–FICA | 23 | 13 | 5 | 36 | 29 | 16 | 9 | 21 | 14 |
| **Fig 7** | NK–PCPS | 23 | 14 | 4.1 | 36 | 61 | 14 | 9 | 19 | 12 |
| **Fig 8** | LSWM (Form 1)–RR | 17 | 15 | 4.8 | 11 | 6.7 | 12 | 5 | 14 | 8 |
| **Fig 8** | LSWM (Form 2)–RR | 17 | 15 | 4.8 | 11 | 7.4 | 12 | 5 | 14 | 8 |
| **Fig 9** | BB–BB | 34 | 13 | 4 | 77 | NA | 19 | 15 | 30 | 24 |
| **Fig 9** | DD–DD | 37 | 14 | 3.8 | 67 | NA | 19 | 18 | 34 | 24 |
| **Fig 9** | NK–NK | 31 | 13 | 3.6 | 76 | NA | 14 | 17 | 28 | 24 |
| **Fig 10** | BB–BB | 34 | 13 | 4 | 70 | NA | 19 | 15 | 30 | 24 |
| **Fig 10** | DD–DD | 37 | 14 | 3.8 | 66 | NA | 19 | 18 | 34 | 24 |
| **Fig 10** | NK–NK | 33 | 14 | 3.8 | 76 | NA | 15 | 18 | 30 | 25 |
| **Fig 11/12** | Entire Figure | 55 | 13 | 4 | 99 | NA | 27 | 28 | 46 | 37 |
| **Fig 13** | Entire Figure | 42 | 14 | 4 | 123 | NA | 20 | 22 | 37 | 28 |

## S8. Self-Selected Engagement by Task Platform

In Study 1, we defined the first session a participant completed of each task (e.g., their first Flanker session, their first List Sorting session) as the target amount for data collection. Any subsequent sessions of that same task were classified as optional repeats. In this Study 1 cohort, 234 of 352 PixelDOPA sessions (66%) were optional, compared to 0 of 112 NIH sessions (0%). In Study 2, two sessions per game was the target amount for data collection. Any subsequent sessions of that same task were classified as optional repeats, and optional repeats accounted for 396 of 726 PixelDOPA sessions (55%). The total number of trials obtained in both studies can be seen in Tables S2 and S3.

## S9. Effect of Trial Count on Results

Because trial counts varied slightly between pD and NIHT task pairs, we capped each task pair at the minimum of the two, with random selection of trials (500 bootstrap draws). With NK–PCPS, DD–DCCS, and BB–FICA, yielding 36, 38, and 29 trials per pair on average, trial-matched correlations remained robust: NK–PCPS $r$ = 0.62 (95% CI 0.56–0.69, $n$ = 23); DD–DCCS $r$ = 0.61 (95% CI 0.58–0.63, $n$ = 23); BB–FICA $r$ = 0.52 (95% CI 0.47–0.57, $n$ = 23).

Furthermore, including all data from all sessions for each participant for all task pairs yielded similar effects: NK–PCPS $r$ = 0.59 ($p = 2.9 \times 10^{-3}$, $n$ = 23); DD–DCCS $r$ = 0.64 ($p = 9.0 \times 10^{-4}$, n = 23); BB–FICA $r$ = 0.40 ($p$ = 0.065, $n$ = 22).

## S10. All-Session Data Quality

Tables 2 and 3 in the main text report data quality from the first valid session (Study 1) or first test-retest pair (Study 2) per participant, matching the analyses in Figures 6–10. The tables below report data quality across all completed sessions, including optional repeats.

All-session exclusion rates were substantially higher than first-session rates for pD tasks (e.g., BB: 26% vs. 8.7% under RT), reflecting the inclusion of unsupervised and repeated sessions where fatigue or inattention was more likely. NIHT exclusion rates also increased modestly (e.g., DCCS: 5.0% vs. 1.2%). Importantly, gRT criteria recovered substantial numbers of trials in the all-session pool (+112 NK, +89 DD, +365 BB in Study 1; +99 NK,

+165 DD, +561 BB in Study 2), a benefit largely absent in the first-session analysis where session quality was already high. This pattern suggests that gRT's primary value for data retention emerges in extended or unsupervised testing, precisely the conditions where pD's scalability advantage is greatest.

**Table S2**: Trial-level data quality across all sessions in Study 1 (convergent validity). Unlike Table 2, which reports only the first valid session per participant, this table includes all optional repeat sessions. Exclusion percentages reflect trials removed under classical RT or gaze-based RT (gRT) criteria across the full session pool. Higher exclusion rates relative to Table 2 reflect the inclusion of unsupervised and repeated sessions where fatigue or inattention was more likely.

| Task | Eligible | Included | Sessions | Trials | RT Excl. % | gRT Excl. % | Trials Salvaged |
| --- | --- | --- | --- | --- | --- | --- | --- |
| PCPS | 25 | 25 | 24 | 1461 | 3.1% | — | — |
| DCCS | 26 | 26 | 25 | 988 | 5.0% | — | — |
| FICA | 25 | 25 | 25 | 736 | 1.9% | — | — |
| NK | 25 | 25 | 69 | 2727 | 14% | 10% | +112 |
| DD | 26 | 26 | 53 | 1878 | 17% | 13% | +89 |
| BB | 25 | 25 | 35 | 1413 | 26% | 8.5% | +365 |

**Table S3:** Trial-level data quality across all sessions in Study 2 (test-retest reliability). Unlike Table 3, which reports only the first test-retest session pair per participant, this table includes all completed sessions. Exclusion rates are higher than in Table 3 because the paired-session analysis pre-selects sessions that passed session-level quality criteria, whereas this table includes sessions that were subsequently excluded.

| Task | Eligible | Included | Sessions | Trials | Sessions | RT Excl. % | gRT Excl. % | Trials Salvaged |
| --- | --- | --- | --- | --- | --- | --- | --- | --- |
| NK | 34 | 34 | 165 | 6177 | 165 | 8.7% | 7.3% | +99 |
| DD | 37 | 37 | 111 | 3652 | 111 | 14% | 10% | +165 |
| BB | 34 | 34 | 96 | 3313 | 96 | 22% | 11% | +561 |